\documentclass[aps,prb,showpacs,letterpaper]{revtex4}
\usepackage{graphics,epsfig}
\usepackage{amssymb,revsymb}
\usepackage{bm}

\begin{document}
\title{Nonlinear effects for island coarsening and stabilization
during strained film heteroepitaxy}
\author{Champika G. Gamage and Zhi-Feng Huang}
\affiliation{Department of Physics and Astronomy, Wayne State
  University, Detroit, Michigan 48201, USA}

\date{\today}
%********************************************************************
\begin{abstract}

Nonlinear evolution of three-dimensional strained islands or quantum 
dots in heteroepitaxial thin films is studied via a continuum elasticity 
model and the development of a nonlinear dynamic equation 
governing the film morphological profile. All three regimes of island 
array evolution are identified and examined, including a film instability 
regime at early stage, a nonlinear coarsening regime at intermediate times, 
and the crossover to a saturated asymptotic state, with detailed behavior
depending on film-substrate misfit strains but not qualitatively on finite 
system sizes. The phenomenon of island stabilization and saturation, 
which corresponds to the formation of steady but non-ordered arrays of
strained quantum dots, occurs at later time for smaller misfit strain.
It is found to be controlled by the strength of film-substrate wetting
interaction which would constrain the valley-to-peak mass transport and
hence the growth of island height, and also determined by the effect of 
elastic interaction between surface islands and the high-order strain 
energy of individual islands at late evolution stage.
%as identified from the time-dependent spatial distribution of elastic
%energy density at the film surface. 
The results are compared to previous experimental and theoretical 
studies on quantum dots coarsening and saturation.

\end{abstract}
%*********************************************************************
\maketitle

%**********************************************************************
\section{Introduction}

The formation of surface nanostructures such as islands or quantum
dots during heteroepitaxy of strained films has attracted continuing great 
interest, due to its importance in both fundamental understanding of
material growth and its wide range of applications particularly for
optoelectronic nano devices
\cite{re:stangl04,re:kiravittaya09,re:berbezier09}.
One of the underlying mechanisms of island formation has been
attributed to the occurrence of morphological instability of the
strained film (i.e., the Asaro-Tiller-Grinfeld instability
\cite{re:asaro72}), for which the competition between the stabilization
effect of film surface energy and the destabilization effect of system
elastic energy due to film-substrate misfit strain plays a key
role. This involves the procedure of stress release in the film via
surface mass transport driven by local energy gradient and
consequently the formation of coherent nonplanar surface structures
like nanoscale islands \cite{re:srolovitz89,re:spencer91}. Compared
to other formation mechanisms such as thermally activated nucleation,
a main feature here is the continuous, nucleationless nature of the
film instability and the subsequent island growth that starts from
rough surface \cite{re:sutter00,re:tromp00,re:tersoff02}. It
results in more regular and correlated arrays of self assembled
quantum dots that are appealing for a variety of applications.

The onset of this type of quantum dots or strained islands formation has
been well understood, as studied via detailed instability analysis for
both single-component \cite{re:srolovitz89,re:spencer91} and
alloy strained films \cite{re:guyer95,re:spencer01,re:huang02,re:desai10} 
as well as multilayers/superlattices \cite{re:shilkrot00,re:huang03}.
Such film instability, showing as surface ripples and cell-like undulations
which are the precursor of coherent island formation,
has been observed in experiments of semiconductor heteroepitaxial films
such as SiGe \cite{re:sutter00,re:tromp00}. On the other
hand, the subsequent nonlinear evolution of these strained dots that are
well beyond the initial linear stability stage, is more complicated
and much less understood. A typical phenomenon is the coarsening of
quantum dot islands, showing as the increase of average island size
during film evolution and the shrinking of small dots. Such scenario
has been observed in experiments of Ge/Si(001)
\cite{re:ross98,re:ribeiro98,re:rastelli05,re:mckay08},
SiGe/Si(001) \cite{re:floro00}, and InAs/GaAs(001)
\cite{re:krzyzewski04}, although with different mechanisms and behavior
of coarsening reported. For the example of Ge-Si systems that have been
extensively studied, nonlinear island coarsening rate that deviates
from classical results of Ostwald ripening has been found; however, it
was associated with different mechanisms of island morphology
evolution in different experiments, such as the effect of island
shape transition accompanying the coarsening process
\cite{re:ross98,re:ribeiro98,re:rastelli05} or a contrasting
kinetic picture that incorporates elastic interactions but not
island shape transition \cite{re:floro00}. Also, the slowing
\cite{re:ross98} or suppression \cite{re:mckay08} of coarsening
process at late stage and the resulting saturated, stabilized quantum
dot arrays \cite{re:dorsch98,re:ribeiro98,re:mckay08} have been
observed in some experiments, but not others \cite{re:floro00}.

The corresponding theoretical/computational study on nonlinear island
evolution is also far from conclusive. Most efforts are based on
continuum approaches including continuum elasticity theory 
\cite{re:yang93,re:spencer93b,re:spencer94,re:chiu99,re:liu01,re:xiang02,re:liu03,%
re:golovin03,re:tekalign04,re:eisenberg05,re:pang06,re:levine07,re:aqua07} 
and phase field methods \cite{re:muller99,re:kassner01}, in addition
to other modeling techniques that incorporate crystalline
details, such as kinetic Monte Carlo method with elastic interaction 
\cite{re:nandpati06,re:schulze09,re:niu11} and the recently developed
phase-field-crystal (PFC) model and the associated amplitude equation
formalism \cite{re:elder02,re:huang08,re:wu09,re:elder10}.
For continuum elasticity modeling, which is the current main avenue
for studying strained island coarsening due to the large length and time
scales involved, much recent focus has been put on the derivation
and simulation of nonlinear evolution equations through approximating
the system elasticity and the dynamics of film morphology via perturbation
methods. Two limits of system configuration were addressed in early
studies of such approach, including the limit of perfectly rigid
substrate \cite{re:spencer93b,re:golovin03} and the case of
infinitely thick strained film \cite{re:xiang02}. More recently,
similar approximation has been applied to heteroepitaxial systems
consisting of a strained thin film grown on an elastic substrate as
configured in most experiments. The corresponding reduced nonlinear
evolution equations have been derived and simulated, based on the
long-wave or small-slope approximation of film surface profile
\cite{re:tekalign04,re:levine07,re:aqua07} or the assumption
of small surface gradient \cite{re:pang06}. Some physical mechanisms 
in thin film growth have also been incorporated, such as the wetting
effect between the strained film and the underlying substrate.
These approaches are more efficient for large scale
simulations, as compared to directly solving the full system
elasticity problem and the corresponding full dynamic equation
of morphological profile, which instead is of high computational cost
and hence usually involves limited system size, island number, and
evolution time particularly for three-dimensional (3D) systems.

Despite the success of these theoretical approaches in describing
properties of quantum dot formation and film morphology, some behavior of
film nonlinear evolution, particularly the process of island coarsening
vs. saturation, is still not well understood, with inconsistent
results given in different studies \cite{re:chiu99,re:liu03,%
re:eisenberg05,re:pang06,re:levine07,re:aqua07,re:aqua10}.
For the case of film annealing as examined in most simulations,
coarsening of strained island arrays has been reproduced, although
with different coarsening rate found in different approaches 
\cite{re:liu03,re:levine07,re:aqua07}. One of the main difference 
in these work is the result for asymptotic and steady
state of the film morphology. Stable arrays of quantum dots that
persist after the coarsening stage have been obtained in both studies
of the reduced nonlinear evolution equation \cite{re:pang06,re:aqua10} 
and the direct solution of the full elasticity problem
\cite{re:chiu99,re:eisenberg05}, consistent with the observation
in some Si-Ge experiments \cite{re:dorsch98,re:ribeiro98,re:mckay08}.
However, as in some other experiments \cite{re:floro00}
such scenario of the suppression or cessation of island coarsening
was not found in other modeling processes
\cite{re:liu03,re:levine07}, and recent nonlinear
analysis of a system evolution equation suggested that a regular
quantum dot array would be unstable as a result of subcritical
bifurcation \cite{re:levine07}. Note that in many previous studies
the saturation of coarsening islands has been attributed to the effect 
of surface energy anisotropy \cite{re:chiu99,re:eisenberg05,re:aqua10},
although the saturation phenomenon has also been observed in recent
simulation without such anisotropy effect \cite{re:pang06}.
These discrepancies in modeling results could probably be related to
different types of approximation and various ways of small variable
expansion and truncation involved in the approaches, and/or the
difficulty in simulating large enough system size and long enough
evolution time required for experimental comparison. 

In this paper we focus on the nonlinear evolution of strained quantum
dot islands grown epitaxially on an elastic substrate, based on a
continuum elasticity model and the development of a systematic
approach for approximately solving the film-substrate elastic state
via a perturbation analysis in Fourier space. Results up to second- 
and third-order perturbation of surface morphology are presented, and
our approach can be readily extended to incorporate higher-order
solutions. We can then derive a new nonlinear evolution equation
governing the dynamics of strained film morphology, which allows us to
systematically examine the detailed behavior of island evolution at
large enough spatial and temporal scales. This nonlinear equation, with
the incorporation of the wetting effect and also a second-order
truncation in the elastic solution, is applied to the study of the
coarsening and saturating process in post-deposited, annealing
films. Our focus is on systems of small misfit strains, which
correspond to large enough length scale of the resulting surface
structure as compared to the scale of crystalline lattice. (This is
based on recent studies \cite{re:huang08} showing that continuum
approaches, such as the continuum elasticity theory developed here,
can well describe the films in weak strain limit, but not for large
misfit stress which would lead to qualitatively different behavior of
surface islands due to the effects of discrete lattice structure.) The
whole series of island evolution can be reproduced in our numerical
simulations, including three characteristic regimes: the development 
of morphological instability and island formation at early times, 
nonlinear coarsening of islands at intermediate stage, and the
slowing of such coarsening process which leads to a saturated state of
steady quantum dot arrays. %The time range of each regime,
%in particular the crossover between coarsening and saturation stages,
%highly depends on the value of film-substrate misfit strain.
We also investigate the mechanisms underlying the phenomenon of island 
stabilization and saturation, based on the study of wetting effect on the 
constraint of surface mass transport between island valleys and peaks, and 
also on a detailed examination of strain relaxation process via studying 
the temporal evolution and spatial distribution of elastic energy density 
at the film surface. We identify a new factor responsible for suppressing 
the island growth and coarsening, which is attributed to the effect of 
high-order elastic energy of individual islands and the elastic 
interaction between them.

%**********************************************************************
\section{Model}

Assume that a strained film of spatially varying height $h(x,y,t)$ is
deposited epitaxially on a semi-infinite elastic substrate that
occupies the region $z<0$. The misfit strain in the film with respect
to the substrate is given by $\epsilon = (a_{f}-a_{s})/a_{s}$, where
$a_f$ and $a_s$ are the lattice spacings of the film and the substrate
respectively. For such coherent, dislocation-free system, the
evolution of the film surface morphological profile $h(x,y,t)$ is
governed by
\begin{equation}
\frac{\partial h}{\partial t}=\Gamma_h \sqrt{g}\nabla^2_s 
\frac{\delta \mathcal F}{\delta h}+v,
\label{eq:h_t}
\end{equation}
where $\Gamma_h$ is the kinetic coefficient determined by surface
diffusion, $\nabla^2_s$ is the surface Laplacian, $v$ is the
deposition rate, and $g=1+|\nabla h|^2$. Here the effect of
film-substrate interdiffusion is neglected. 
The total free energy functional $\mathcal F$ consists of two parts,
including the elastic energy $\mathcal F_{el} = \int_{-\infty}^{h}
d^3r {\mathcal E}$ where ${\mathcal E}$ represents the strain energy
density, and the surface free energy
$\mathcal F_s = \int d^2r \gamma_s(h) \sqrt{g}$ where $\gamma_s$ is the
thickness-dependent, isotropic surface tension with the effect of
wetting interaction between the film and substrate incorporated. The
dynamic equation (\ref{eq:h_t}) can then be rewritten as 
\cite{re:spencer91,re:tekalign04,re:levine07}
\begin{equation}
\frac{\partial h}{\partial t}=\Gamma_h \sqrt{g}\nabla^2_s
\left [ \gamma \kappa + W(h) + {\mathcal E}^f \right ]+v,
\label{eq:e}
\end{equation}
where $\kappa$ is the mean surface curvature, $\gamma$ is the surface
energy density, ${\mathcal E}^f$ gives
the film elastic energy density at the surface $z=h$, and the wetting
potential $W$ can be approximated via a phenomenological glued-layer
wetting model \cite{re:levine07} 
\begin{equation}
W(h)=-w \left ( \frac{h}{h_{ml}} \right )^{-\alpha_w}
e^{-h/h_{ml}}.
\label{eq:wetting}
\end{equation}
Here $w$ gives the strength of the film-substrate wetting interaction,
$h_{ml}$ is the characteristic wetting-layer thickness that is
usually of few monolayers, and the exponent $\alpha_w$ ($>0$) gives
the singularity of the potential $W$ in the limit of $h \rightarrow 0$
that emulates the persistence of the wetting layer during film
evolution \cite{re:levine07}.

The formulation of elasticity for this film-substrate system has been
well established \cite{re:spencer91}. In isotropic, linear
elasticity theory (i.e., with harmonic approximation), the elastic
energy density is given by ${\mathcal E} = \frac{1}{2} \sigma_{ij}
u_{ij}$, where $i,j=x,y,z$, and $u_{ij}$ is the linear elastic strain
tensor defined by $u_{ij} = (\partial_j u_i + \partial_i u_j)/2$
(with $u_i$ the displacement field). From Hooke's law for isotropic
elastic system, the stress tensor $\sigma_{ij}$ in the strained film
is expressed by 
\begin{equation}
\sigma_{ij} = 2 \mu \left [ \frac{\nu}{1-2 \nu} \delta_{ij}
u_{kk}+u_{ij}-\frac{1+\nu}{1-2\nu}\epsilon\delta_{ij} \right ],
\label{eq:sigmF}
\end{equation}
where $\mu$ is the shear modulus and $\nu$ is the Poisson ratio. The
stress tensor in the substrate is also given by Eq. (\ref{eq:sigmF})
with $\epsilon=0$. Here for simplicity we have assumed equal elastic
constants in the film and substrate, which corresponds to the
situation in most experimental systems where the difference of elastic
constants between film and substrate materials is not significant.

Since the elastic relaxation occurs at a time scale of orders of
magnitude faster than that of the atomic diffusion process and the
associated system morphological evolution, it is usually
assumed that the mechanical equilibrium condition $\partial_j
\sigma_{ij}=0$ is always satisfied in both film and substrate. 
Using Eq. (\ref{eq:sigmF}) we can obtain Navier's equations
in the whole film-substrate system
\begin{equation}
(1-2\nu)\partial^2_j u_i + \partial_i \partial_j u_j = 0.
\label{eq:equilibrium}
\end{equation}
The corresponding boundary condition on the top film surface is given by
\begin{equation}
\sigma^f_{ij} n_j=0 \qquad \textrm{at } z=h(x,y,t),
\label{eq:a}
\end{equation}
due to the neglecting of external pressure on the free surface,
while the boundary conditions at the film-substrate interface
is determined by the continuity of stress and displacement fields:
\begin{equation}
\sigma^f_{ij}n_j=\sigma^s_{ij}n_j \quad \textrm{and} \quad
u^f_i=u^s_i \qquad \textrm{at } z=0.
\label{eq:c}
\end{equation}
Here $n_j$ is the unit vector normal to the film surface or the
film-substrate interface, and the subscripts ``\textit{f}'' and
``\textit{s}'' refer to the film and substrate phases, respectively.
Also, inside the substrate region which is far away from the film we have
\begin{equation}
u^s_{i}, u^s_{ij} \rightarrow 0 \qquad \textrm{for } 
z \rightarrow -\infty.
\label{eq:d}
\end{equation}
 
%************************************************************************
\section{Perturbation analysis and nonlinear evolution equation}
\label{sec:method}

To solve this elasticity problem, we adopt a perturbation analysis in
Fourier space based on the expansion of the small vertical variation
of film surface profile. More specifically, given the Fourier
transform of the film morphological profile 
\begin{equation}
h=\bar{h}+\sum_{\bm q}\hat{h}(\mathbf{q},t)e^{i(q_x x+q_y y)},
\label{eq:h_hat}
\end{equation}
where $\bar{h}=h_0+vt$ is the average film thickness at any time $t$
(with $h_0$ the initial film thickness), the Fourier components of
the displacement field $\hat{u}_i(\mathbf{q})$, stress tensor
$\hat{\sigma}_{ij}(\mathbf{q})$ ($i,j = x,y,z$), and film elastic
energy density $\hat{\mathcal E}^f$ are expanded in the
order of surface perturbation $\hat{h}(\mathbf{q})$, i.e.,
\begin{eqnarray}
& u_i = \bar{u}_i + \sum_{\bm q}\hat{u}_i(\mathbf{q})e^{i(q_x x+q_y y)},& 
\qquad \hat{u}_i = \hat{u}_i^{(1)} + \hat{u}_i^{(2)} + \hat{u}_i^{(3)}
+ \cdots, \nonumber\\
& \sigma_{ij} = \bar{\sigma}_{ij }+ \sum_{\bm q}\hat{\sigma}_{ij} 
(\mathbf{q})e^{i(q_x x+q_y y)},& \qquad
\hat{\sigma}_{ij} = \hat{\sigma}_{ij}^{(1)} + \hat{\sigma}_{ij}^{(2)}
+ \hat{\sigma}_{ij}^{(3)} + \cdots, \label{eq:u_sigma_hat}\\
& \mathcal{E}^f = \bar{\mathcal E}^f + \sum_{\bm q} \hat{\mathcal E}^f
(\mathbf{q})e^{i(q_x x+q_y y)},& \qquad 
\hat{\mathcal E}^f = \hat{\mathcal E}^{(1)f} + \hat{\mathcal E}^{(2)f}
+ \hat{\mathcal E}^{(3)f} + \cdots. \nonumber
\end{eqnarray}
For the 0th-order base state with planar, uniformly strained film,
the elasticity solutions are given by \cite{re:spencer91}:
$\bar{u}^f_i=0$ ($i=x,y$) and
$\bar{u}^f_z=\bar{u}^f_{zz}z$ for the displacement fields, the strain
tensor $\bar{u}^f_{ij}=0$ except for $\bar{u}^f_{zz}=\epsilon
(1+\nu)/(1-\nu)$, the stress tensor $\bar{\sigma}^f_{ij}=0$ except
for $\bar{\sigma}^f_{xx}=\bar{\sigma}^f_{yy} =-2\mu u^0_{zz}$, and the
0th-order elastic energy density $\bar{\mathcal E}^f=E \epsilon^2 /
(1-\nu)$ (where $E$ is the Young's modulus). For the
substrate, the corresponding base state is stress-free, with
$\bar{u}_i^s = \bar{u}^s_{ij} = \bar{\sigma}^s_{ij} = 0$.

The elastic properties at higher orders can be obtained by
substituting the expansions (\ref{eq:h_hat}) and
(\ref{eq:u_sigma_hat}) into Eqs. (\ref{eq:equilibrium})--(\ref{eq:d}).
In Fourier space the Navier's equations (\ref{eq:equilibrium}) can be
rewritten as
\begin{eqnarray}
&&(1-2\nu)(\partial_z^2-q^2)\hat{u}_j^{(\xi)}+iq_j \left [
  iq_x\hat{u}_x^{(\xi)} +iq_y\hat{u}_y^{(\xi)} 
+\partial_z\hat{u}_z^{(\xi)} \right ] =0,  \qquad \textrm{for }j=x,y,
\label{eq:q} \\
&&(1-2\nu)(\partial_z^2-q^2)\hat{u}_z^{(\xi)}+\partial_z \left [
  iq_x\hat{u}_x^{(\xi)} +iq_y\hat{u}_y^{(\xi)}
+\partial_z\hat{u}_z^{(\xi)} \right ] =0, \label{eq:s}
\end{eqnarray}
for $\xi$th order expansion ($\xi=1,2,3,...$).
The corresponding general solutions have the same format as that
obtained in Ref. \onlinecite{re:spencer91} for 1st order equations,
which read
\begin{equation}
\hat{u}_i^{(\xi)f}=\left[\begin{array}{c}
\alpha_x^{(\xi)} \\
\alpha_y^{(\xi)} \\
\alpha_z^{(\xi)} \end{array}\right]\cosh(qz)+\left[\begin{array}{c}
\beta_x^{(\xi)}\\
\beta_y^{(\xi)}\\
\beta_z^{(\xi)}\end{array}\right]\sinh(qz)-\left[\begin{array}{c}
C^{(\xi)}iq_x/q\\
C^{(\xi)}iq_y/q\\
D^{(\xi)}\end{array}\right]z\sinh(qz)-\left[\begin{array}{c}
D^{(\xi)}iq_x/q\\
D^{(\xi)}iq_y/q\\
C^{(\xi)}\end{array}\right] z\cosh(qz)
\label{eq:uf}
\end{equation}
for the film, and
\begin{equation}
\hat{u}_i^{(\xi)s}=\left[\begin{array}{c}
\alpha_x^{(\xi)}\\
\alpha_y^{(\xi)}\\
\alpha_z^{(\xi)}\end{array}\right]e^{qz}-\left[\begin{array}{c}
iq_x/q\\
iq_y/q\\1\end{array}\right]B^{(\xi)}ze^{qz}
\label{eq:us}
\end{equation}
for the substrate after using the boundary conditions (\ref{eq:c}) and
(\ref{eq:d}) at the film-substrate interface and inside the substrate.
The coefficients $\alpha_i^{(\xi)}$, $\beta_i^{(\xi)}$, $C^{(\xi)}$,
$D^{(\xi)}$, and $B^{(\xi)}$ in Eqs. (\ref{eq:uf}) and (\ref{eq:us}) are
determined via the expansion of boundary conditions
(\ref{eq:a})--(\ref{eq:c}) in orders of perturbation $\hat{h}$.
Note that the 1st order solution has been known with the use of
linearized boundary conditions \cite{re:spencer91,re:guyer95},
with the perturbed elastic energy density being given by
\begin{equation}
\hat{\mathcal E}^{(1)f} = -\frac{2E(1+\nu)}{1-\nu} \epsilon^2 q
\hat{h}(\mathbf{q}).
\label{eq:E1}
\end{equation}

For the 2nd order expansion of the boundary conditions, at
the top surface of the film, $z=h$, Eq. (\ref{eq:a}) can be written as
\begin{equation}
-\sum_{\mathbf{q}'} i(q_x-q'_x)\hat{\sigma}^{(1)f}_{jx}(\mathbf{q}')
\hat{h}(\mathbf{q}-\mathbf{q}') -\sum_{\mathbf{q}'} i(q_y-q'_y)
\hat{\sigma}^{(1)f}_{jy}(\mathbf{q}') \hat{h}(\mathbf{q}-\mathbf{q}')
+\hat{\sigma}^{(2)f}_{jz}(\mathbf{q})=0,
\label{eq:j}
\end{equation}  
while the continuity of stress at the film-substrate interface $z=0$
(i.e., Eq. (\ref{eq:c})) yields
\begin{equation}
\hat{\sigma}^{(2)f}_{jz}(\mathbf{q})=\hat{\sigma}^{(2)s}_{jz}(\mathbf{q}),
\label{eq:m}
\end{equation}
with $j=x,y,z$.
Substituting Eqs. (\ref{eq:uf}) and (\ref{eq:us}) to these boundary
conditions (\ref{eq:j}) and (\ref{eq:m}), the second order
coefficients of the solution can be obtained as follows:
\begin{eqnarray}
&&q\alpha_z^{(2)}=q\beta_z^{(2)}=-e^{-q\bar{h}} \left [
  \frac{a_1^{(2)}q_x+b_1^{(2)}q_y}{2\mu q}(1-2\nu+q\bar{h}) 
  -\frac{c_1^{(2)}}{2\mu}(2-2\nu+q\bar{h})\right ],
\label{eq:az} \nonumber\\ 
&&iq_x\alpha_x^{(2)}+iq_y\alpha_y^{(2)}=iq_x\beta_x^{(2)}
+iq_y\beta_y^{(2)} =e^{-q\bar{h}} \left [
  \frac{a_1^{(2)}q_x+b_1^{(2)}q_y}{2\mu q}
  (q\bar{h}-2+2\nu)+\frac{c_1^{(2)}}{2\mu}(1-2\nu+q\bar{h}) \right ],
\label{eq:axy} \nonumber\\
&&C^{(2)}=D^{(2)}=B^{(2)}=e^{-q\bar{h}} \left [
  -\frac{a_1^{(2)}q_x+b_1^{(2)}q_y}{2\mu q}+\frac{c_1^{(2)}}{2\mu} \right ],
\label{eq:CDB}
\end{eqnarray}
where
\begin{eqnarray}
& a_1^{(2)}q_x+b_1^{(2)}q_y &= \sum_{\mathbf{q}'} \hat{h}(\mathbf{q}-\mathbf{q'})
\hat{h}(\mathbf{q}') \left \{\frac{2E\epsilon}{q'(1-\nu)}
\left [ q_x(q_x-q_x')(q_x'^2+\nu q_y'^2) + q_y(q_y-q_y')(q_y'^2+\nu
  q_x'^2) \right ] \right. \nonumber \\
 && \left. +2E\epsilon \frac{q_x'q_y'}{q'} \left [
     q_x(q_y-q_y')+q_y(q_x-q_x') \right] \right \},
\label{eq:a1b1}
\end{eqnarray}
and
\begin{equation}
c_1^{(2)} = \sum_{\mathbf{q}'} \hat{h}(\mathbf{q}-\mathbf{q'})
\hat{h}(\mathbf{q}') \frac{E\epsilon}{1-\nu} \left [
  q_x'(q_x-q_x')+q_y'(q_y-q_y') \right ].
\label{eq:c1}
\end{equation}
Based on the above solution, we can determine the second order elastic
energy density, i.e.,
\begin{eqnarray}
& \hat{\mathcal E}^{(2)f} &=\sum_{\mathbf{q}'} \left [\frac{1+\nu}{2E}
  \hat{\sigma}_{ij}^{(1)f}(\mathbf{q}') \hat{\sigma}_{ij}^{(1)f}
  (\mathbf{q}-\mathbf{q}') -\frac{\nu}{2E}
  \hat{\sigma}_{ll}^{(1)f}(\mathbf{q})
  \hat{\sigma}_{ll}^{(1)f}(\mathbf{q}-\mathbf{q}') \right ]
\nonumber\\ 
&& +\frac{E\epsilon}{1-\nu} \left [(1-\nu)\frac{a_1^{(2)}q_x+b_1^{(2)}q_y} 
  {\mu q} - (1-2\nu)\frac{c_1^{(2)}}{2\mu} \right ],
\label{eq:second}
\end{eqnarray}
where the expressions of 1st-order stress tensor
$\hat{\sigma}_{ij}^{(1)f}$ at the top surface are given in the
appendix [see Eqs. (\ref{eq:sigxx1})--(\ref{eq:sigyz1})]. The
2nd-order stress tensor can be also calculated, with results shown in 
Eqs. (\ref{eq:sigxx2})--(\ref{eq:sigyz2}) of the appendix.

We then derive the nonlinear evolution equation for film surface
morphology from Eq. (\ref{eq:e}), using the results of perturbed
elastic energy density given in Eqs. (\ref{eq:E1}) and
(\ref{eq:second}). All the terms in the dynamic equation (\ref{eq:e})
are expanded up to second order of surface perturbation $\hat{h}$,
except for $W(h)$ for which the full nonlinear wetting potential form
Eq. (\ref{eq:wetting}) is used. For the surface energy term 
$\gamma \kappa$, noting that the surface curvature $\kappa =
-\bm{\nabla} \cdot [\bm{\nabla} h / \sqrt{1+|\bm{\nabla} h|^2}]$,
we have $\gamma \kappa \sim -\gamma \nabla^2 h + \mathcal{O}(\hat{h}^3)$
and hence only need to keep the linear order term in the 2nd-order
approximation considered here.
To further simplify the calculation, we choose
a length scale $l=\gamma/\mathcal{E}_0$ and a time scale
$\tau=l^4/{\gamma\Gamma_h}$, where the characteristic strain energy
density $\mathcal{E}_0 = 2E\epsilon_0^2 (1+\nu)/(1-\nu)$ with
$\epsilon_0$ a reference misfit value. The resulting
nondimensional dynamic equation for the perturbed surface profile
$\hat{h}(\mathbf{q},t)$ is given by
\begin{equation}
\frac{\partial\hat{h}}{\partial t}=(-q^4+{\epsilon^*}^2q^3)
\hat{h} -q^2\mathcal{W}_{\mathbf{q}}
- {\epsilon^*}^2\sum_{\mathbf{q}'} \hat{h}(\mathbf{q}')
\hat{h}(\mathbf{q}-\mathbf{q}') \Lambda(\mathbf{q},\mathbf{q}'),
\label{eq:height}
\end{equation}
where $\mathcal{W}_{\mathbf{q}}$ is the Fourier transform of the 
rescaled wetting potential $W(h)/\mathcal{E}_0$, 
$\epsilon^*=\epsilon/\epsilon_0$, and
\begin{eqnarray}
&\Lambda(\mathbf{q},\mathbf{q}')&=q^2 \left [ (1-\nu)
  \frac{[\mathbf{q}' \cdot (\mathbf{q}-\mathbf{q}')]^2}
  {q'|\mathbf{q}-\mathbf{q}'|} 
  -\mathbf{q}' \cdot (\mathbf{q}-\mathbf{q}')
  +\nu q'|\mathbf{q}-\mathbf{q}'| \right ] \nonumber \\
&&+\frac{2q}{q'} \left \{ q_x(q_x-q_x')(q_x'^2+\nu q_y'^2)
  +q_y(q_y-q_y')(q_y'^2+\nu q_x'^2)+(1-\nu)q_x'q_y' \left [
    q_x(q_y-q_y')+q_y(q_x-q_x') \right ] \right \}.
\label{eq:lambda}
\end{eqnarray}
In Eq. (\ref{eq:height}), the first term of the right-hand-side is the 
combination of the surface energy contribution and the 1st order elastic 
energy density $\hat{\mathcal{E}}^{(1)f}$, consistent with the previous 
linear-order results \cite{re:spencer91,re:tekalign04,re:levine07}. The 
last term is from the 2nd-order perturbation result of the elastic 
energy density, i.e., Eq. (\ref{eq:second}) for
$\hat{\mathcal{E}}^{(2)f}$. It would be straightforward, although with
more complicated processes, to extend the above approach to incorporate 
higher-order contributions, based on the perturbed analysis
of system elasticity given in Eqs. (\ref{eq:h_hat})--(\ref{eq:us}).
That is, to obtain the $n$th-order elastic results, we can first express
the $n$th-order expansion of the boundary conditions in terms of
$\xi$th-order ($\xi=1,2,...,n-1,n$) stress tensors [similar to the
expression in Eq. (\ref{eq:j})], and use it to calculate the $n$th-order
solution of the displacement field given in Eq. (\ref{eq:uf}); The
corresponding film elastic properties can then be derived,
particularly the $n$th-order elastic energy density
$\hat{\mathcal{E}}^{(n)f}$ at the film surface which can be expressed
as a function of $\xi$th-order ($\xi=1,2,...,n-1$) elastic quantities
that are already known. We have applied this recursive method to
third-order calculations, with results of elastic energy density
$\hat{\mathcal{E}}^{(3)f}$ shown in the appendix. The corresponding
higher-order evolution equation can be obtained via adding
term $-q^2 \hat{\mathcal{E}}^{(3)f}$ to Eq. (\ref{eq:height}).

\section{Linear stability analysis}

In the following studies of film evolution and strained
island dynamics, we focus on Eq. (\ref{eq:height}) with 2nd-order
elastic properties. Here we first perform a linear stability analysis
of Eq. (\ref{eq:height}) to determine the conditions of morphological
instability of the system; such conditions are needed for the nonlinear
calculations given in Sec. \ref{sec:wetting} and Sec. \ref{sec:numeric}. 
Following the standard procedure, we assume an exponential growth
$\hat{h}=\hat{h}_0\exp(\sigma_h t)$ at early time, and apply it to the
linearized evolution equation of $\hat{h}$. The characteristic
equation for the perturbation growth rate $\sigma_h$ is then given by
\begin{equation}
\sigma_h=-q^4+q^3{\epsilon^*}^2-q^2a, 
\label{eq:dispersion}
\end{equation}
where $a=(w^*/h_{ml}^*) (x+\alpha_w) x^{-\alpha_w-1}e^{-x}$,
$x=\bar{h}/h_{ml}$, $w^*=w/\mathcal{E}_0$, and $h_{ml}^*=h_{ml}/l$.
From the above dispersion relation we can identify the condition for
the occurrence of film morphological instability, which is given by
${\epsilon^*}^4 \geq 4a$ or equivalently,
\begin{equation}
\frac{e^x x^{\alpha_w+1}}{x+\alpha_w} \geq 
\frac{4w^*}{{\epsilon^*}^4h_{ml}^*}.
\label{eq:inst}
\end{equation}
The corresponding characteristic wave number of film instability (for
the fastest instability growth mode) can be written as
\begin{equation}
q_{\rm max}= \frac{3}{8} \left [ {\epsilon^*}^2 + \sqrt{{\epsilon^*}^4
  - \frac{32}{9} a} \right ].
\label{eq:qmax}
\end{equation}
Eq. (\ref{eq:inst}) is used to identify the
parameters in our numerical simulations shown below, for which
the initial film instability and hence the appearance of nonplanar
surface morphology or islands are required. Note that for given
film conditions such as misfit strain $\epsilon$ and wetting
parameters, Eq. (\ref{eq:inst}) indicates that due to the
film-substrate wetting effect (with $\alpha_w>0$), the morphological
instability and surface nanostructures will develop only for thick
enough films, with the critical thickness $\bar{h}_c$ determined by
Eq. (\ref{eq:inst}), i.e.,
$e^{x_c} x_c^{\alpha_w+1} / (x_c+\alpha_w) = 4w^*/{\epsilon^*}^4h_{ml}^*$ 
(where $x_c=\bar{h}_c/h_{ml}$); the value of $\bar{h}_c$ increases
with smaller film-substrate misfit strain. Also, the characteristic
size (or wavelength $\lambda = 2\pi / q_{\rm max}$) of
surface structures at the initial stage will decrease with the
increasing average film thickness $\bar{h}$, as can be obtained from
Eq. (\ref{eq:qmax}).

\section{Effects of wetting potential}
\label{sec:wetting}

To validate our model system and the nonlinear dynamic equation 
(\ref{eq:height}) derived above, we first examine the effect of wetting 
potential on film evolution and compare it to the well-known results 
of cusp formation obtained from previous full elasticity calculations 
\cite{re:yang93,re:spencer94}. Since Eq. (\ref{eq:height}) is already 
presented in Fourier space, in our numerical simulations we directly use 
the spectral method with periodic boundary conditions along the 
lateral $x$ and $y$ directions, and also an exponential propagation 
algorithm for time integration \cite{re:cross94}. This allows us to use 
large enough time steps (up to $\Delta t=1$ in most results shown below,
except for the study of groove/cusp formation for which $\Delta t=0.01$ 
is used). For rescaling parameters that are associated with the nondimensional 
equation (\ref{eq:height}), the reference misfit $\epsilon_0=3 \%$ is 
chosen, and thus the length scale can be estimated as $l \simeq 5.5$~nm 
if using the material parameters of Ge/Si system. For simplicity, in this 
work we focus on the case of nongrowing films with deposition rate $v=0$ 
and simulate the annealing process of film evolution. 

\begin{figure}
\centerline{
\includegraphics[width=\textwidth]{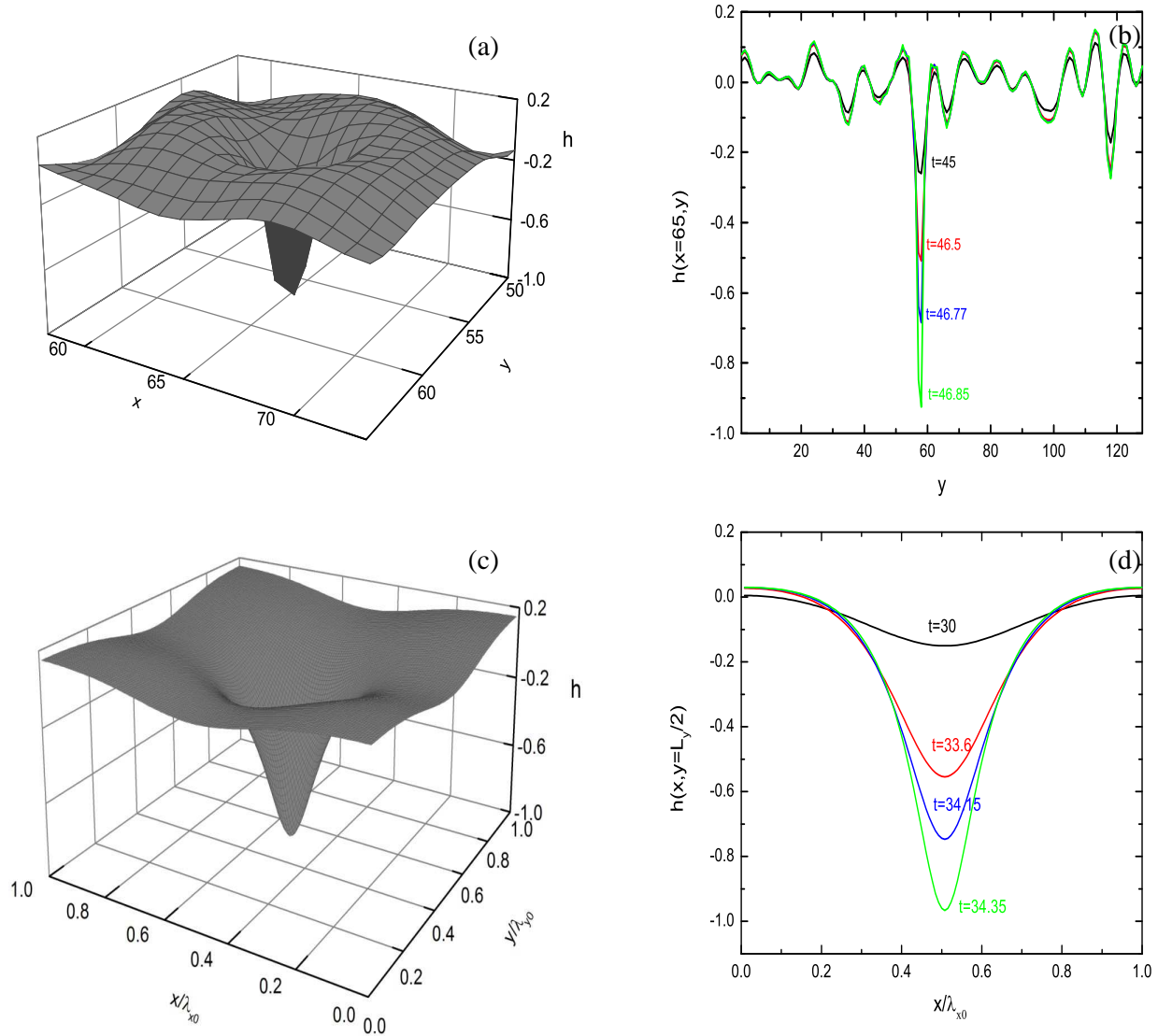}}
\vspace{-0.2in}
\caption{Surface morphologies of 3\% strained films, as obtained from
numerical simulations without the wetting effect. The simulations start either
from small random initial perturbation of a planar film [(a) and (b)] or 
from a doubly-periodic surface profile with wavevector $q_{x0}=q_{y0}=3/4\sqrt{2}$ 
and amplitude $A_0=0.01$ [(c) and (d)]. Both 3D morphologies, (a) at
$t=46.77$ for a portion of a $128 \times 128$ system and (c) at $t=34.35$ 
for system size $\lambda_{x0} \times \lambda_{y0}$, and also time evolution 
of 2D cross-section profiles are shown.}
\label{fig:cusps}
\end{figure}

As found first by Yang and Srolovitz \cite{re:yang93} and later in various 
numerical studies of 2D \cite{re:spencer94,re:xiang02} and 3D 
\cite{re:golovin03,re:pang06} systems via solving either the full elasticity 
problem or the reduced nonlinear evolution equations, deep grooves or cusps 
will form in stressed solid systems without the incorporation of wetting 
effect. This is well reproduced in our numerical results of Eq. (\ref{eq:height}), 
as shown in Fig. \ref{fig:cusps} for 3\% misfit films with $\nu=1/3$ and
initial film thickness $h_0=0$. Two types of initial conditions are used: 
(1) a small random disturbance of a planar film of thickness $h_0$, with results 
of a $128 \times 128$ system presented in Fig. \ref{fig:cusps} (a) and (b), and 
(2) a weakly perturbed film with doubly-periodic sinusoidal surface profile 
$h = h_0 + A_0 [\cos (q_{x0} x) + \cos (q_{y0} y)]$, as given in panels 
(c) and (d) of Fig. \ref{fig:cusps} for a $\lambda_{x0} \times \lambda_{y0}$
system (where $\lambda_{x0} = 2\pi / q_{x0}$ and $\lambda_{y0} = 2\pi / q_{y0}$). 
For condition (2) the initial perturbed amplitude is set as $A_0=0.01$, and 
a perturbed wavevector $q_{x0}=q_{y0}=3/4\sqrt{2}$ is chosen, corresponding 
to the wave number of the most linearly unstable mode determined by 
Eq. (\ref{eq:qmax}). In both cases the formation of singular cusps or deep
grooves and their rapid growth have been found during the film evolution, 
as evidenced by the 3D morphological profiles given in Fig. \ref{fig:cusps} 
(a) and (c), and also from the results of time-evolving 2D cross-section profiles 
shown in (b) and (d) which are consistent with the previous 2D results of Spencer
and Meiron \cite{re:spencer94} and Xiang and E \cite{re:xiang02}.

\begin{figure}
\centerline{
\includegraphics[height=3.in]{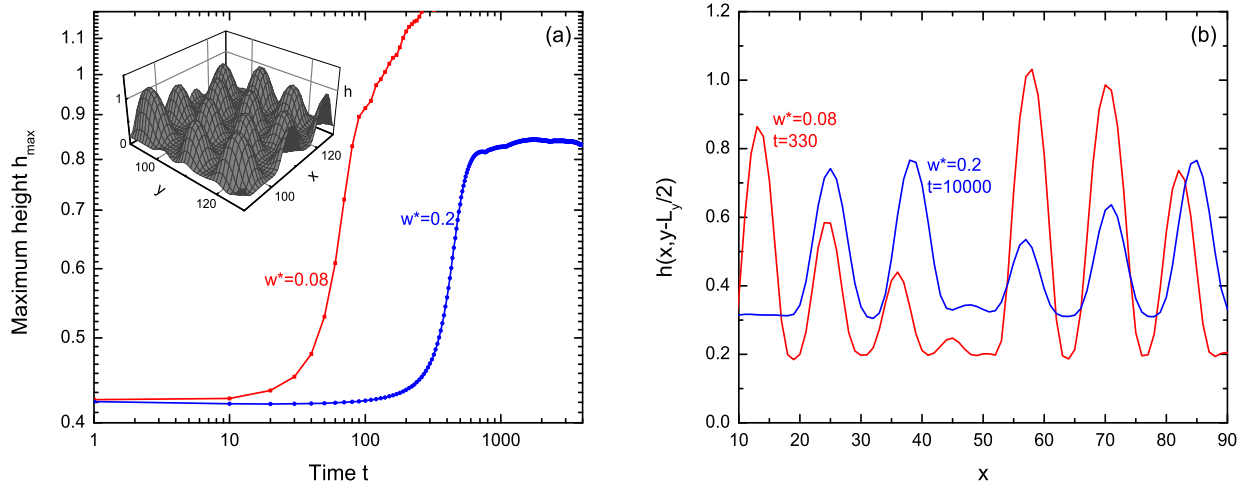}}
\caption{Time evolution of 3\% strained films with different wetting
strength $w^*=0.08$ and $0.2$. (a) Evolution of maximum surface height,
with a 3D island morphology for $w^*=0.08$ at $t=330$ shown in the inset;
(b) 2D cross-section profiles at $y=L_y/2$ for $w^*=0.08$ at $t=330$ and 
$w^*=0.2$ at $t=10000$.}
\label{fig:wetting_w}
\end{figure}

To incorporate the wetting effect, in our calculations we use a
pseudospectral method; that is, we first evaluate
the wetting potential $W(h)$ from Eq. (\ref{eq:wetting}) in real space
and then obtain its Fourier component $\mathcal{W}_{\mathbf{q}}$ as used
in the dynamic equation (\ref{eq:height}). As expected, the cusp/groove
singularity is completely suppressed by the film-substrate wetting
interaction, and arrays of strained islands or quantum dots will form
and grow. This has been verified by our numerical results shown in 
Fig. \ref{fig:wetting_w}, where we have used the parameters of
$\epsilon=3\%$, $h_0=0.41$, $h_{ml}^*=0.3$, $\alpha_w=2$, 
$w^*=0.08$ or $0.2$, and simulation time step $\Delta t=1$.
However, another type of growth instability would occur when the strength 
of wetting interaction is not strong enough (e.g., $w^*=0.08$ in 
Fig. \ref{fig:wetting_w}), showing as the rapid increase of island heights 
beyond initial time stage and then the blow-up of numerical solution at
late times. Such instability with unbound growth of island height is
absent for stronger wetting effect, such as the effect of $w^*=0.2$ shown
in Fig. \ref{fig:wetting_w}(a) which gives the stabilization and saturation
of island evolution. This can be understood from the fact that the wetting 
interaction tends to prevent the depletion of the film-substrate wetting 
layer and hence suppress the mass transport from the valley of an island 
to its top, leading to the constraint of island height as a result of mass
conservation. Such effect would increase with the strength of wetting
interaction, as can be seen from the results given in 
Fig. \ref{fig:wetting_w}(b): thicker film layers between surface islands
and shallower valleys are found for larger wetting strength $w^*$, as a 
result of stronger suppression on the valley-to-peak diffusion process.

%**********************************************************************
\section{Results of nonlinear evolution}
\label{sec:numeric}

\begin{figure}
\centerline{
\includegraphics[height=3.9in]{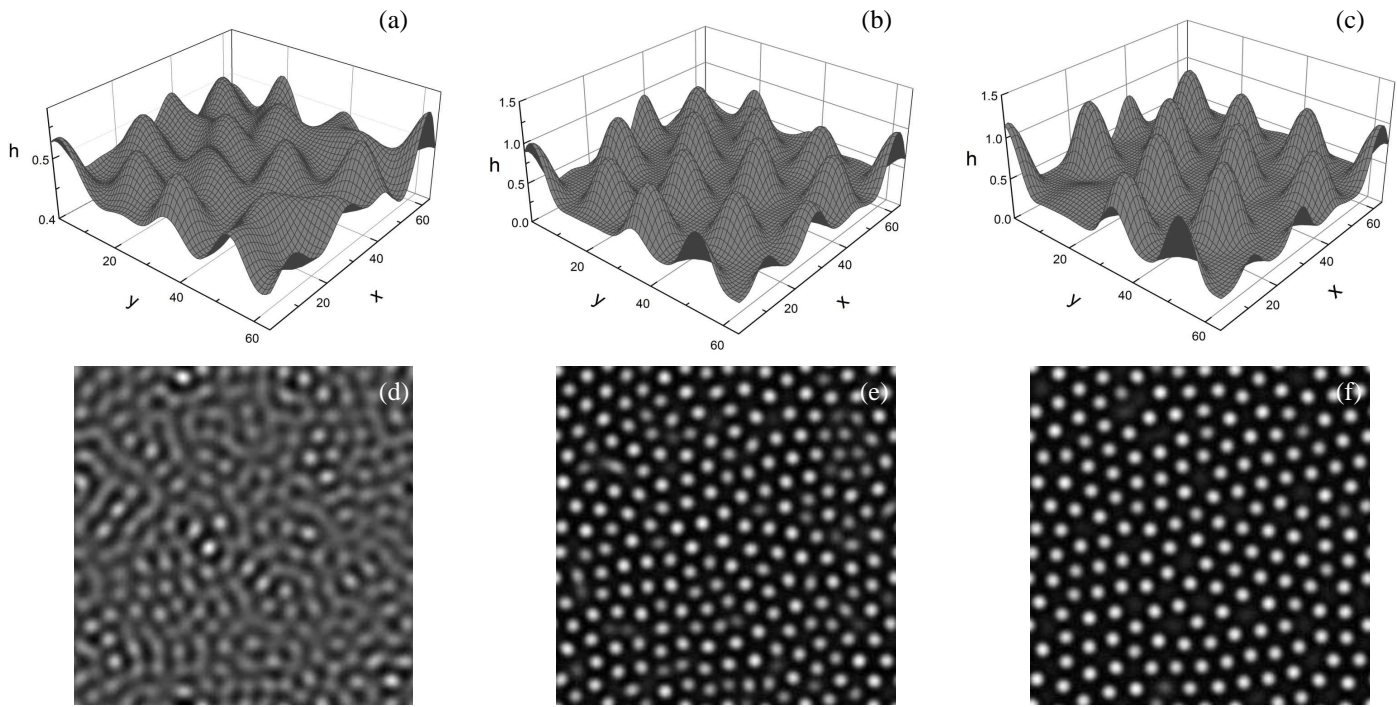}}
\caption{Morphological profiles of a $2.5\%$ strained film,
  at simulation times $t=1000$ [(a) and (d)], 2000 [(b) and (e)], 
  and 10000 [(c) and (f)]. A
  fraction of 3D morphologies in a $256 \times 256$ system is given in
  panels (a)--(c), while the corresponding 2D gray scale top-view
  images of the full system size are shown in (d)--(f).}
\label{fig:h2}
\end{figure}

To examine the detailed evolution of strained film morphology more 
systematically, we have conducted numerical simulations of the full dynamic
equation (\ref{eq:height}) for different small film-substrate misfit strains
$\epsilon=2\%$, $2.5\%$, and $3\%$ (in such weak strain limit the continuum 
elasticity approach can be well applied, as shown in most recent studies 
\cite{re:huang08}). The parameters for the wetting potential are chosen as 
$h_{ml}^*=0.3$, $\alpha_w=2$, and $w^*=0.2$, with all other parameters
the same as those given in Sec. \ref{sec:wetting}. In our simulations we 
have used 3 different system sizes $L_x \times L_y$ for each 
parameter set, including the lateral dimensions of $128 \times 128$,
$256 \times 256$, and $512 \times 512$, to examine any possible
artifacts of finite size effects. A numerical grid spacing 
$\Delta x = \Delta y =1$ is adopted, and the integration time step
is chosen as $\Delta t=1$. The quantitative results given below
have been averaged over 20 independent runs for system sizes $128
\times 128$ and $256 \times 256$, and 10 runs for $512 \times 512$.
Also, each simulation starts with a rescaled initial film
thickness of $h_0=0.67$ for misfit strain $\epsilon=2\%$, $h_0=0.5$
for $\epsilon=2.5\%$, and $h_0=0.41$ for $\epsilon=3\%$, all of which
are within the corresponding instability parameter region for each
misfit as determined by Eq. (\ref{eq:inst}).

\begin{figure}
\centerline{
\includegraphics[height=3.9in]{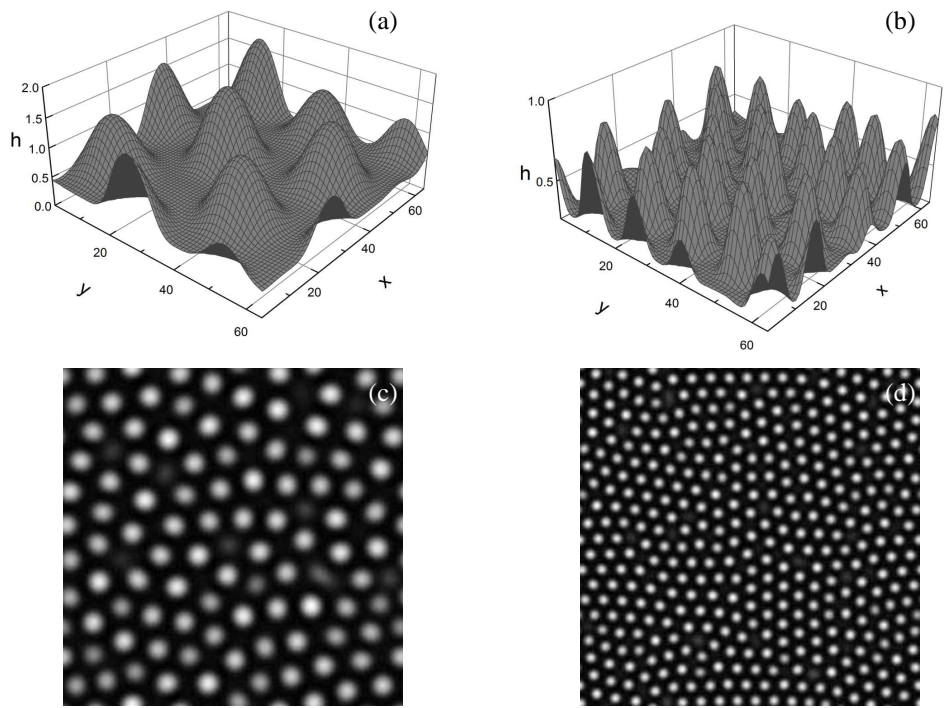}}
\caption{Morphological profiles of (a) $2\%$ and (b) $3\%$ strained
  films at late time stage of $t=10000$. Only a portion of the $256
  \times 256$ system is shown in the 3D images of (a) and (b). The
  corresponding 2D top-view images of the full system size are given
  in (c) and (d).}
\label{fig:h}
\end{figure}

Typical simulation results of film evolution and the formation and
dynamics of quantum dot arrays are illustrated in Fig. \ref{fig:h2},
for $2.5\%$ mismatch between the film and substrate. At the beginning
stage surface undulations occur due to film morphological instability
determined in Eq. (\ref{eq:inst}), leading to the formation of
strained surface islands or quantum dots as shown in
Fig. \ref{fig:h2}(a). Note that at different surface locations, islands
will form and grow gradually at different rates due to the nonlinear
effects of elastic interaction. Island coarsening occurs at
the next stage, showing as the growth of some quantum dots at the
expense of other shrinking ones and hence the decrease of island
density on the film surface. This can be seen more clearly in the
corresponding 2D top-view images of Figs. \ref{fig:h2}(d)--(f),
which give the comparison of island distribution between times
$t=1000$, $2000$, and $10^4$. Such coarsening process becomes
much slower as time increases, and the system would approach an
asymptotic state with steady arrays of strained quantum dots.
As expected, this late-time state of film surface structures 
highly depends on the value of film-substrate misfit strain, with an
increase of island density and a decrease of island spacing for
larger misfits. This has been confirmed in our results of $2\%$
and $3\%$ films given in Fig. \ref{fig:h}, as compared to the $2.5\%$
film shown in Figs. \ref{fig:h2}(c) and \ref{fig:h2}(f). In our
simulations no long-range spatial order can be found for quantum dot
arrays, even at the late-time stage, agreeing with the observation
of most experimental and theoretical studies.

\begin{figure}
\centerline{
\includegraphics[height=2.2in]{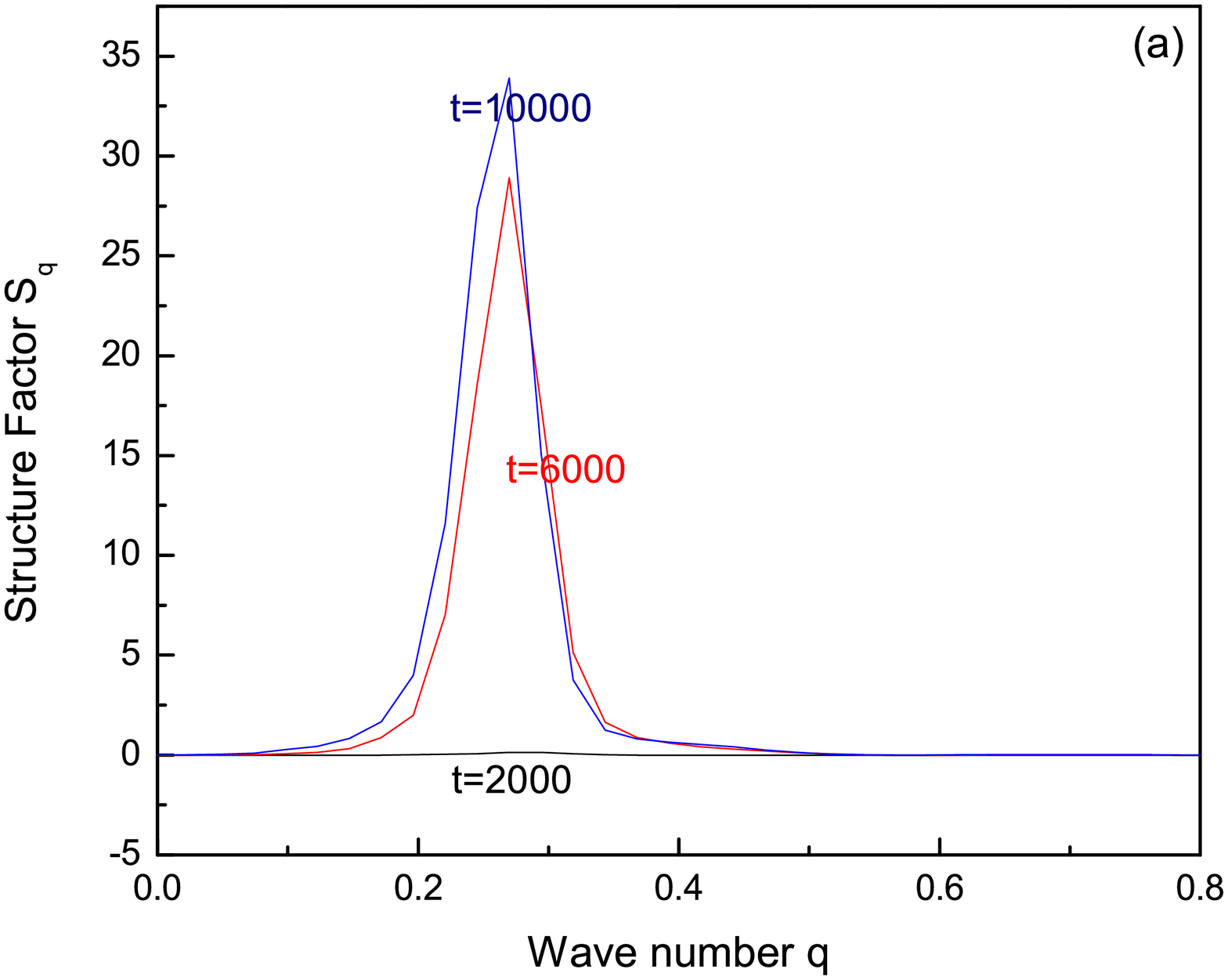} \hskip -40pt
\includegraphics[height=2.2in]{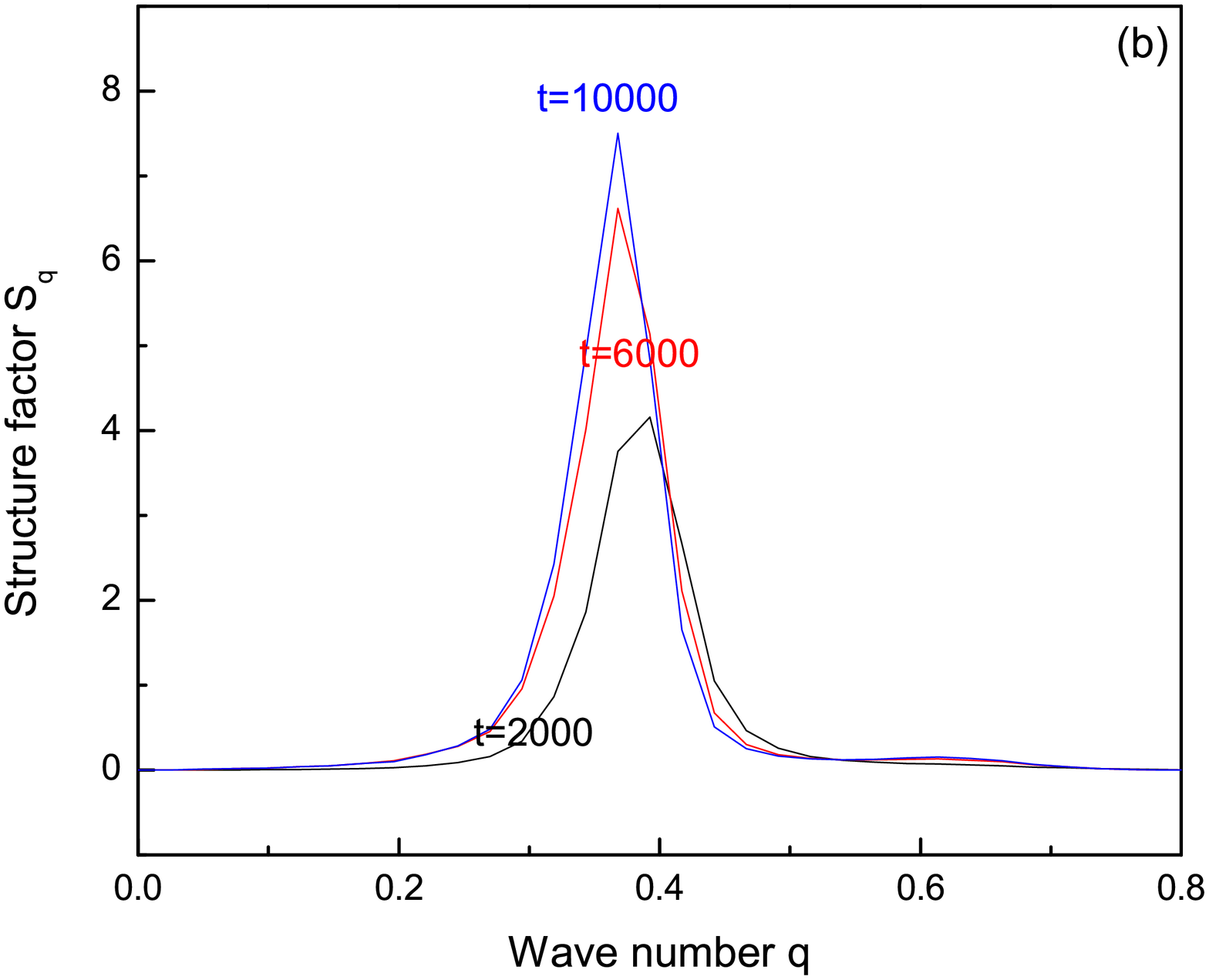} \hskip -40pt
\includegraphics[height=2.2in]{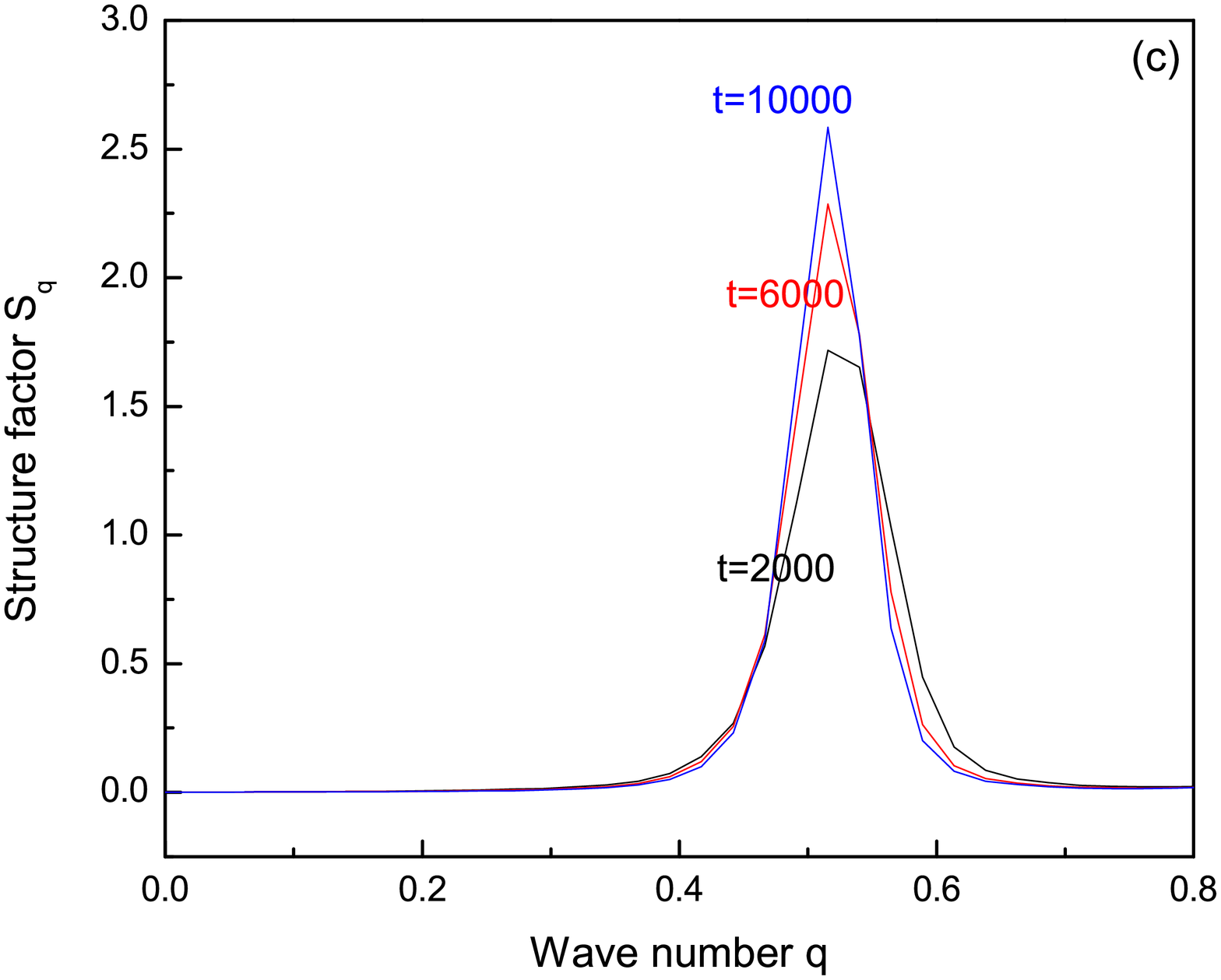}}
\caption{Structure factor of the surface height as a function of wave
  number $q$, for different misfits (a) $\epsilon=2\%$, (b)
  $\epsilon=2.5\%$, and (c) $\epsilon=3\%$, system size $256 \times
  256$, and times $t=2000$, $6000$, and $10000$.}
\label{fig:Sq}
\end{figure}

\begin{figure}
\centerline{
\includegraphics[height=3.2in]{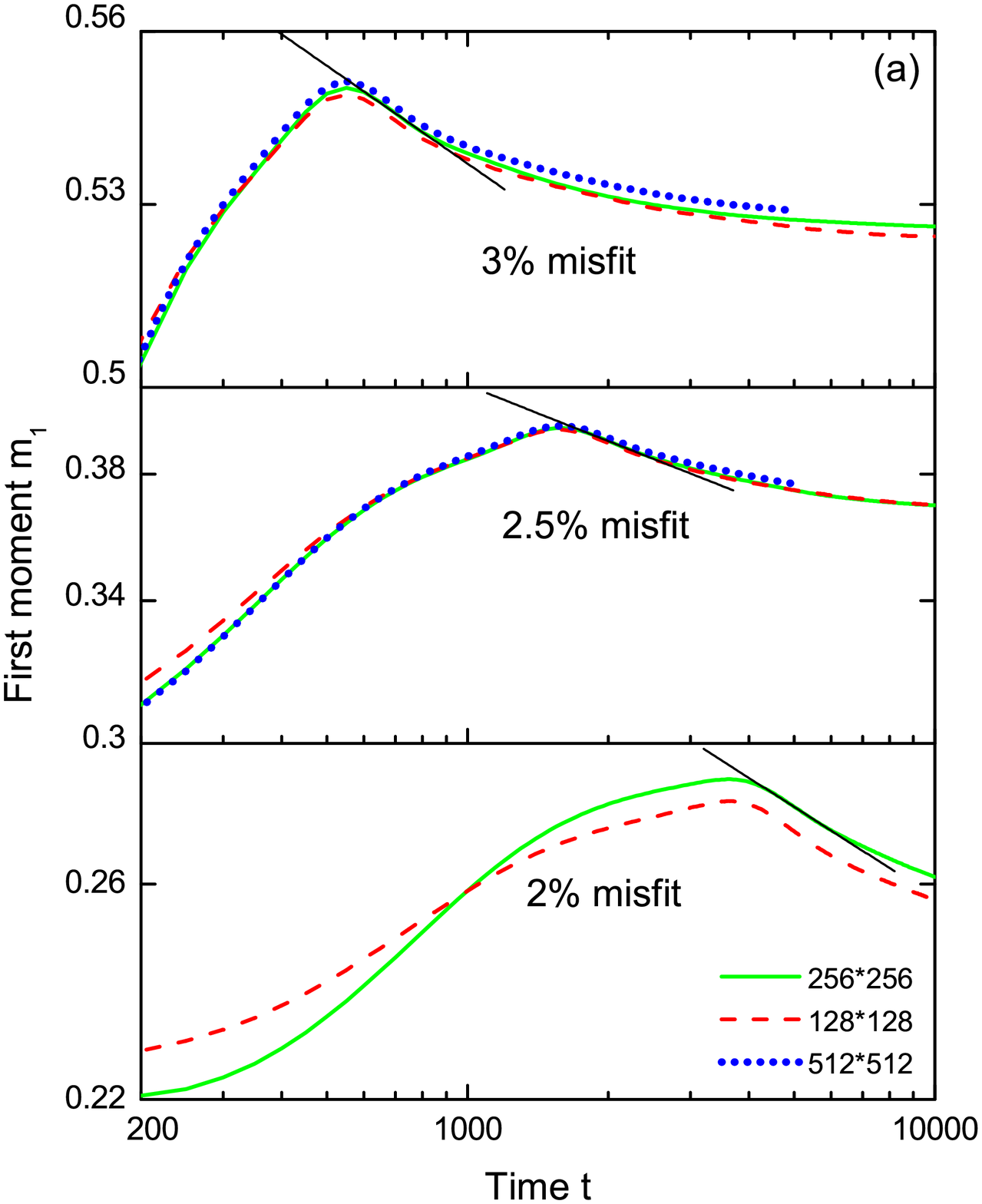} \hskip -10pt
\includegraphics[height=3.2in]{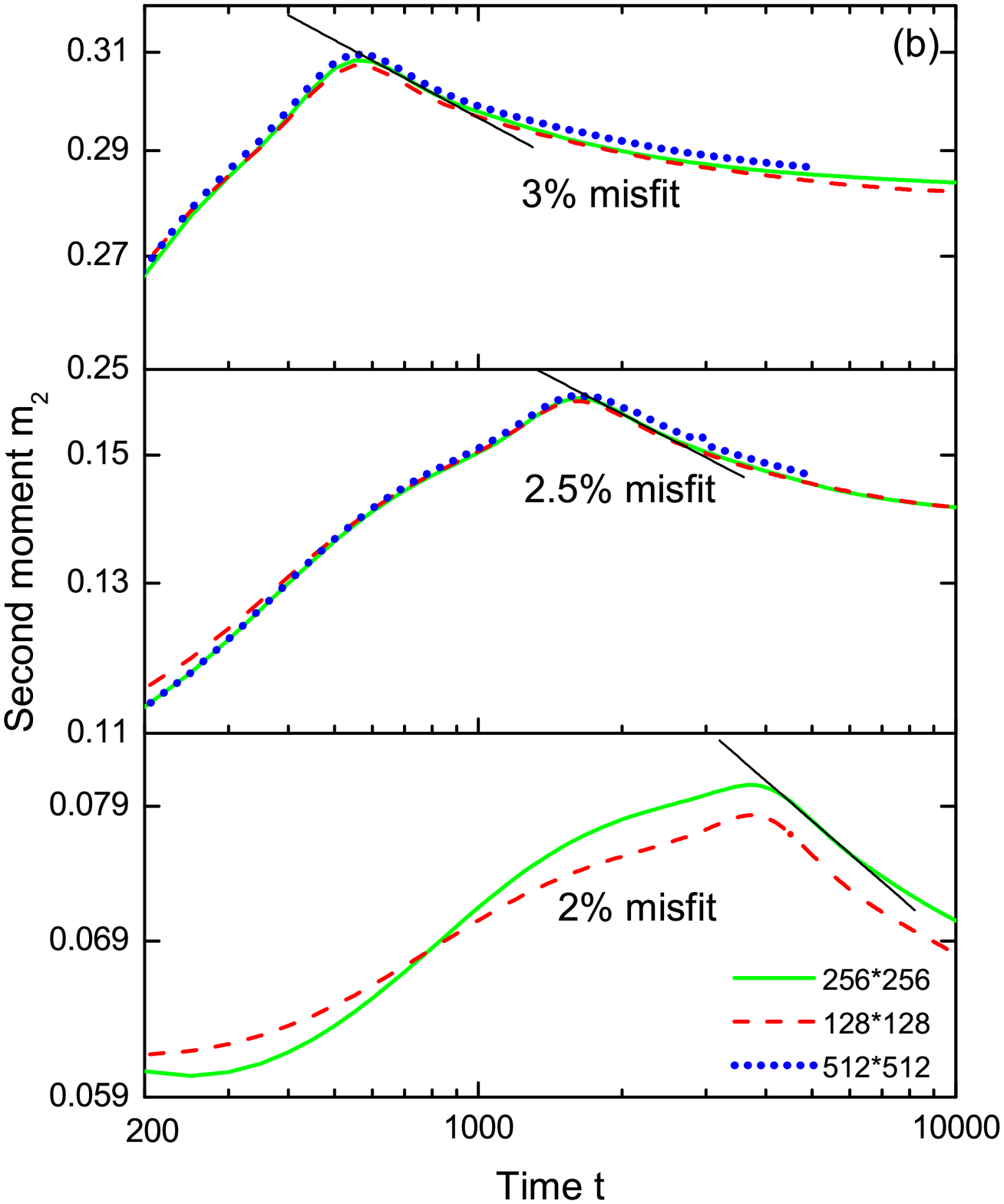} \hskip -10pt
\includegraphics[height=3.2in]{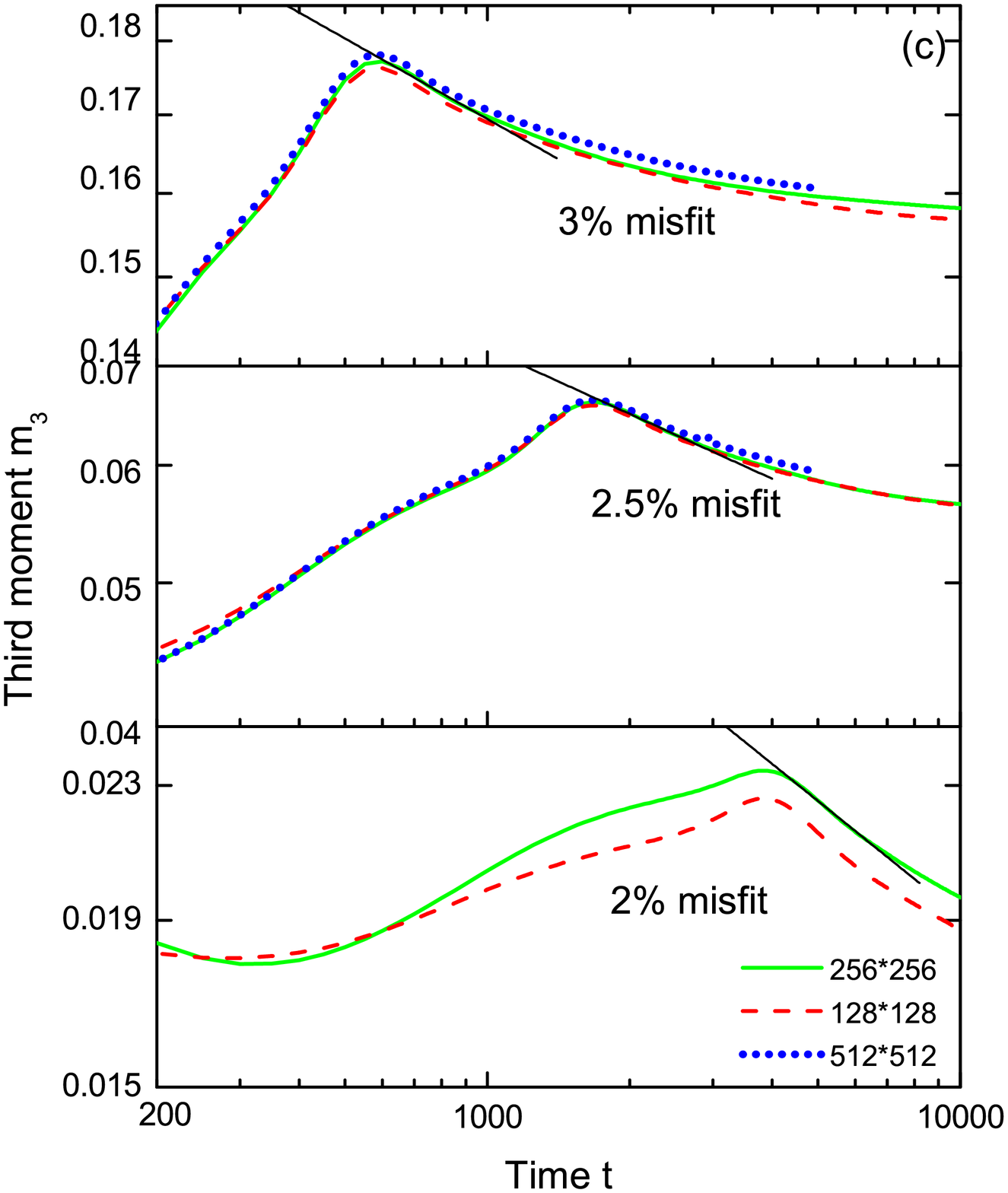}}
\caption{Time evolution of the three moments of structure factor: (a)
  $m_1$, (b) $m_2$, and (c) $m_3$, for misfit strains $\epsilon=2\%$,
  $2.5\%$, and $3\%$. The solid lines represent the simulation results
  of grid size $256 \times 256$, whereas the dashed and dotted lines
  represent the results for grid sizes $128 \times 128$ and $512
  \times 512$ respectively. Power law fittings (i.e., the thin
  lines) at the beginning of coarsening stage for the $256 \times 256$
  system are also shown.}
\label{fig:mn}
\end{figure}

\begin{figure}
\centerline{
\includegraphics[height=3.3in]{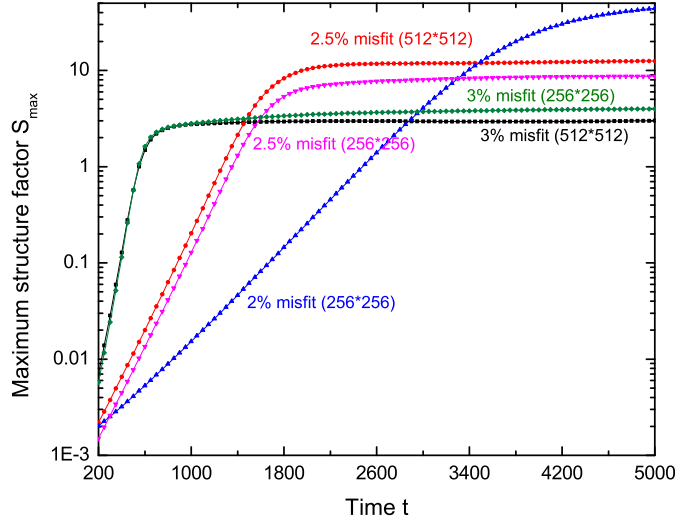}}
\caption{Time evolution of the maximum value of structure factor,
  for misfits $\epsilon=2\%$, $2.5\%$, and $3\%$. Results for
  different system sizes $256 \times 256$ and $512 \times 512$ are
  shown for comparison. }
\label{fig:Sqmax}
\end{figure}

\begin{figure}
\centerline{
\includegraphics[height=3.3in]{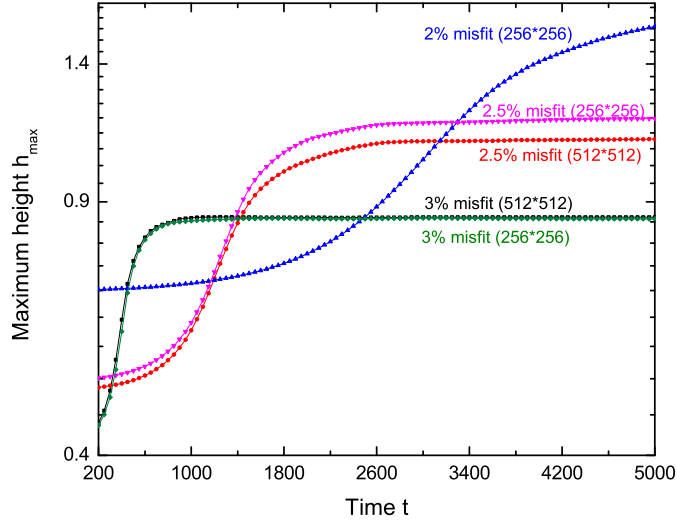}}
\caption{Time evolution of maximum surface height for misfits
  $\epsilon=2\%$, $2.5\%$, and $3\%$ and system sizes $256 \times 256$
  and $512 \times 512$.}
\label{fig:hmax}
\end{figure}

To quantify the above results, we have analyzed the film surface
morphology through various time-dependent parameters, including the
structure factor of surface height, its moments, the maximum
height of surface profile, and the surface roughness, as shown in
Figs. \ref{fig:Sq}--\ref{fig:hmax} with results of different
system sizes presented. The structure factor is defined as
$S(q,t)=\langle |\hat{h}(\mathbf{q},t)|^2 \rangle_{\hat{\bm q}}$, with a
circular average over orientation $\hat{\bm q}$ of the wave number. 
Typical results of $S(q,t)$ are given in Fig. \ref{fig:Sq}, for
different misfit strains of $2\%$, $2.5\%$, and $3\%$. As the misfit
increases, the wave number related to the peak location of the
structure factor becomes larger, corresponding to smaller island
spacing and also higher quantum dot density as already seen in
Figs. \ref{fig:h}(c), \ref{fig:h2}(f), and \ref{fig:h}(d). Also,
for smaller misfit strain larger time is needed for the initial
formation of islands (if we compare the $t=2000$ curves in
Figs. \ref{fig:Sq}(a)--(c)), consistent with the result of $\sigma_h$
in the linear stability analysis.

Details of the time evolution of quantum dot islands can be
characterized by the calculation of various moments of the structure
factor. The $n$th moment of $S(q,t)$ is defined as
\begin{equation}
m_n(t) = \frac{\int dq q^n S(q,t)}{\int dq S(q,t)},
\end{equation}
which yields the information of the characteristic size and spatial
scale of surface structures \cite{re:sagui94}. We have calculated
the first three moments of $S(q,t)$, with time evolution results for
three different misfits given in Figs. \ref{fig:mn}(a) (for $m_1$),
\ref{fig:mn}(b) (for $m_2$), and \ref{fig:mn}(c) (for $m_3$).
Three characteristic regimes of film evolution can be identified in
each simulation: the process of surface instability development and
island formation at early times, coarsening of these strained islands
at intermediate stage, and an asymptotic stage of island saturation
at late times. The first two stages can be distinguished clearly in
the results of moments shown in Fig. \ref{fig:mn}, as separated by the
turning point (i.e., maximum of $m_n$) of the time evolution curve for
each moment. The increase of $m_n$ at the first time
stage is due to the continuous appearance of new islands at various
times as observed in our simulations, and thus the decrease of average
island spacing. The time range of this early stage of island formation
is longer for smaller misfit strain, as a result of overall smaller
instability growth rate [see Eqs. (\ref{eq:dispersion})--(\ref{eq:qmax})].
This has been verified in Fig. \ref{fig:mn}, via comparing the three
panels of misfits $3\%$, $2.5\%$, and $2\%$ (from top to bottom) for
results of each moment.

Once most islands have formed (i.e., when the moments $m_n$ reach
maximum values), they start to coarsen so that the average distance
between islands increases, leading to the decrease of $m_n$.
A power-law behavior of coarsening, $m_n(t) \sim t^{-\beta_n}$, can be
obtained, but such behavior is limited to a transient time range 
at the beginning of the coarsening stage. As shown in
Fig. \ref{fig:mn}, this time range is smaller for larger
film-substrate misfit strain $\epsilon$, with faster crossover to a
saturated state. Also, slower coarsening rate has been
found for larger misfit, corresponding to smaller coarsening exponents
$\beta_n$ which are identified as (for system size $256 \times 256$): 
For $\epsilon=2\%$, %in the time range 4200-5550 
$\beta_1 = 0.1010 \pm 0.0008$, $\beta_2 = 0.181 \pm 0.002$, and
$\beta_3 = 0.235 \pm 0.003$; 
For $\epsilon=2.5\%$, %in the time range 1750-2300
$\beta_1 = 0.0702 \pm 0.0007$, $\beta_2 = 0.120 \pm 0.002$, and 
$\beta_3 = 0.145 \pm 0.004$;
For $\epsilon=3\%$, %between times 600-800, 
$\beta_1 = 0.0449 \pm 0.0009$, $\beta_2 = 0.076 \pm 0.003$, and 
$\beta_3 = 0.090 \pm 0.006$. Note that 
if the structure factor is assumed to obey a simple dynamic scaling
behavior due to coarsening, one would usually expect that $m_n(t) \sim
t^{-n\beta_1}$; i.e., $\beta_n = n\beta_1$. However, the above results
of coarsening exponents in the intermediate time range for
all different misfit strains do not support this assumption, and we
cannot identify a simple format of scaling for the structure
factor. This might be attributed to the complex relaxation of strain
energy in the film and the nonlinear elastic interaction between
surface islands [see e.g., Eq. (\ref{eq:lambda})] which are more
complicated than that revealed by simple scaling.

\begin{figure}
\centerline{
\includegraphics[height=3.3in]{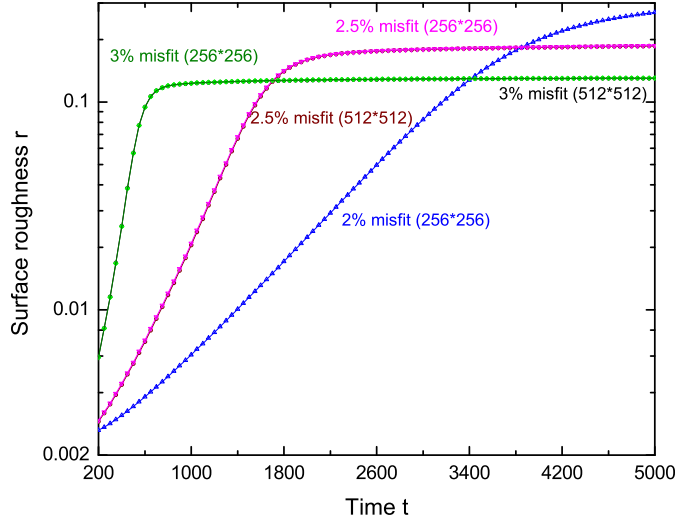}}
\caption{Time evolution of surface roughness for various misfits
  $\epsilon=2\%$, $2.5\%$ and $3\%$. Note that for each misfit,
  results of different system sizes $256 \times 256$
  and $512 \times 512$ almost overlap with each other.}
\label{fig:roughness}
\end{figure}

These two regimes of island formation and coarsening are also
illustrated in our numerical results for the maximum value of the
structure factor $S_{\rm max}$ (Fig. \ref{fig:Sqmax}), the maximum
surface height $h_{\rm max}$ (Fig. \ref{fig:hmax}), and the surface
roughness $r(t) = \langle (h-\bar{h})^2 \rangle ^{1/2}$
(Fig. \ref{fig:roughness}). The growth of all three quantities can
be observed during the first stage of instability growth and island
formation, which corresponds to the same initial time range as the
$m_n$ results shown in Fig. \ref{fig:mn}. Both $S_{\rm max}$ and the
roughness $r(t)$ grow exponentially with time at this stage,
consistent with the behavior of linear instability analyzed
in Sec. \ref{sec:method}. However, at later times during the
coarsening process, these quantities show rather slow growth and
an approach to saturation, even for $h_{\rm max}$. Such phenomenon of
surface roughness saturation has been obtained in a recent study of a
nonlinear evolution equation \cite{re:pang06}, but not in other
studies using different evolution equations (which instead observed
power-law growth) \cite{re:levine07,re:aqua07}. The limited growth of 
maximum surface height given in Fig. \ref{fig:hmax} during island
coarsening has not been reported in those previous studies, which
usually showed a faster growth of $h_{\rm max}$ such as a power-law
behavior \cite{re:levine07}.

Our simulation results also indicate a crossover from the island
coarsening regime to an asymptotic state of steady quantum dot arrays,
showing as saturated values of $S_{\rm max}$, $h_{\rm max}$, and
$r(t)$ (see the $2.5\%$ and $3\%$ results in
Figs. \ref{fig:Sqmax}--\ref{fig:roughness}), and more clearly, the
saturation of $m_n$ given in Fig. \ref{fig:mn}. Such crossover
can be identified through the slowing of the $m_n$ decay after the
transient of power-law-type coarsening, and occurs earlier for
larger misfit strain. To exclude any artifacts from finite size effects,
we have tested our simulations for different system sizes ranging from
$128 \times 128$ to $512 \times 512$, with qualitatively similar
results obtained. This phenomenon of the slowing and saturation of
coarsening process has been found in some experiments of
Ge/Si systems \cite{re:ross98,re:dorsch98,re:ribeiro98,re:mckay08}
and also in some modeling and simulation results based on either direct
solution of system elasticity \cite{re:chiu99,re:eisenberg05}
or reduced film evolution equations \cite{re:pang06,re:aqua10}; 
however, no sign of coarsening termination in annealing films has 
been shown in some other experimental \cite{re:floro00} and 
theoretical \cite{re:liu03,re:levine07,re:aqua07} work.
Also, the effect of different misfit strains on island coarsening
and saturation, which is shown important from our results here, has
not been addressed in most previous studies. Our results given above
suggest that much longer times are needed to observe the slowing or
cessation of island coarsening for smaller misfits, which could be
useful for addressing the discrepancy of experimental observation:
E.g., the suppression or saturation of quantum dot growth can be found
at relatively short annealing times for Ge/Si(001) system with large
misfit ($\sim 4\%$) \cite{re:ross98,re:ribeiro98,re:mckay08},
while it is more difficult to observe in SiGe/Si(001) experiments with
weak misfit strain ($< 1\%$) \cite{re:floro00}.

\section{Discussion}

To understand the mechanisms underlying the saturation phenomenon
given above, it would be helpful to examine the effects of all
contributed terms in the evolution equation (\ref{eq:height}).
As already discussed in Sec. \ref{sec:wetting}, the wetting
potential plays a crucial role on the stabilization of
surface morphology via limiting the valley-to-peak mass transport
and thus the growth of island height. Similar mechanism is expected
during island coarsening, and the stabilization effect of wetting
potential should be also important for the island size saturation.
On the other hand, the dynamics of coarsening involves the 
redistribution of mass between different islands,
%from small and shrinking islands to the large and growing ones,
a process that cannot be constrained by the wetting effect as long
as the wetting layer is not depleted. Thus additional factor(s) must
be in play to account for the island saturation and stabilization
process.

\begin{figure}
\centerline{
\includegraphics[height=3.in]{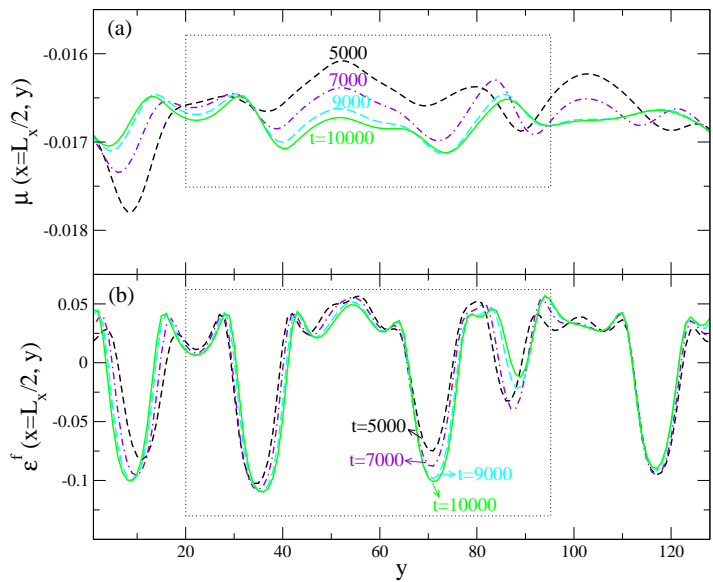}}
\caption{Cross-section profiles of (a) surface chemical potential
  $\mu$ and (b) elastic energy density ${\cal E}^f$, for misfit
  $\epsilon=2.5\%$ and different times $t=5000$, $7000$, $9000$, and
  $10000$. The boxed region will be further studied in
  Fig. \ref{fig:muy_hy}.}
\label{fig:muy}
\end{figure}

\begin{figure}
\centerline{
\includegraphics[height=4.3in]{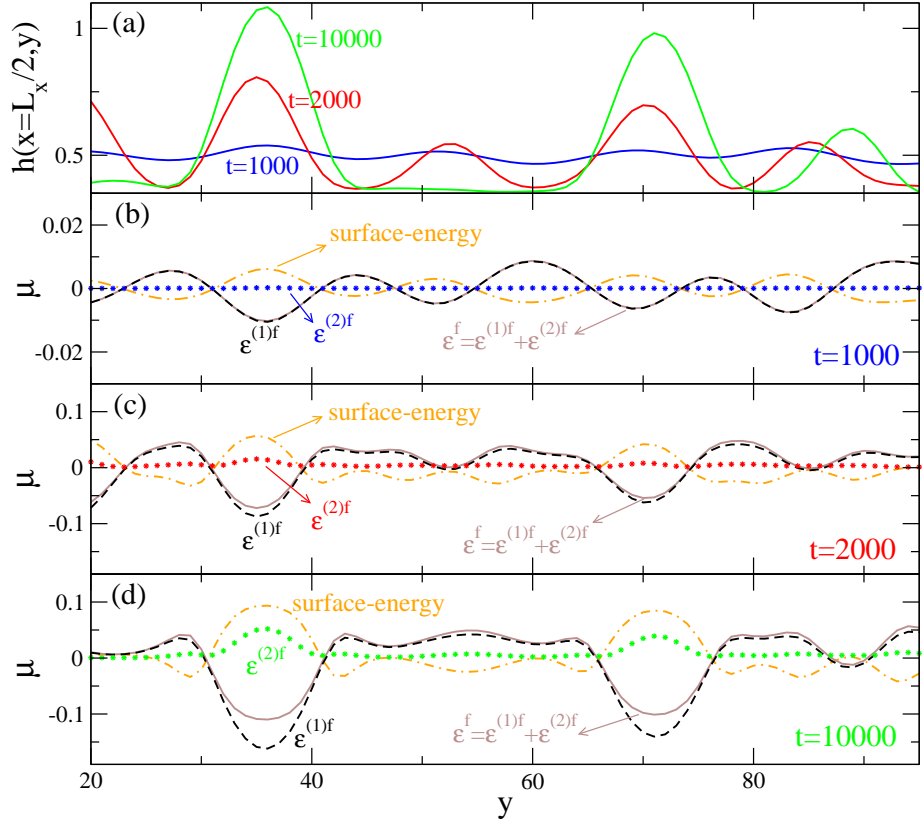}}
\caption{Cross-section profiles of (a) surface morphology and (b)-(d)
  various energy densities, for $2.5\%$ misfit and times $t=1000$ (at
  early stage of instability and island formation), $2000$ (island
  coarsening stage), and $10000$ (saturating stage). Different
  components of the film surface chemical potential are shown,
  including the surface-energy contribution (dot-dashed orange
  curves), 1st-order elastic energy density ${\cal E}^{(1)f}$ (dashed
  black), 2nd-order elastic density ${\cal E}^{(2)f}$ (green stars),
  and the total elastic contribution ${\cal E}^f={\cal E}^{(1)f}+{\cal
    E}^{(2)f}$ (solid brown).}
\label{fig:muy_hy}
\end{figure}

To further illustrate the saturating process, we show in
Fig. \ref{fig:muy} the time-varying profiles of the chemical potential
$\mu=\gamma \kappa + W + {\cal E}^f$ at the film surface and also its
corresponding elastic contribution ${\cal E}^f$ (i.e., the surface
elastic energy density). For simplicity, only the cross-section
results at $x=L_x/2$ are presented, for an example of $2.5\%$ misfit
film. Smaller spatial variations of chemical potential $\mu$ along the 
film surface are obtained at later times [note the very small vertical 
scale in Fig. \ref{fig:muy}(a)], indicating an approach to an asymptotic
saturated state. This is consistent with the results given in
Figs. \ref{fig:mn}--\ref{fig:roughness} for various morphological
properties, and also with the evolution profiles of elastic energy
density ${\cal E}^f$ given in Fig. \ref{fig:muy}(b). Furthermore, in
Fig. \ref{fig:muy_hy} we examine the detailed mechanisms of such
saturation through identifying the time evolution of various
components of chemical potential, including the surface-energy
contribution $\gamma \kappa$ and the first and second order elastic
energy densities ${\cal E}^{(1)f}$ and ${\cal E}^{(2)f}$. We focus on
a small region of 4 islands [see Fig. \ref{fig:muy_hy}(a)],
representing 3 scenarios of quantum dot evolution: (1) large islands
that are growing and saturating, (2) small islands that are
shrinking [see the middle island in Fig. \ref{fig:muy_hy}(a)], and
(3) islands that are migrating (see the one at the right corner). As
expected from previous analysis \cite{re:spencer91}, for an undulated
surface (i.e., in the region of surface islands), the strain energy is 
concentrated at surface valleys but released at peaks; the resulting 
surface elastic energy density gradient would drive the diffusion process 
from the valleys to peaks and thus the growth of surface islands. On the
other hand, this morphological destabilization process is competed
by the stabilization effect of surface energy, showing as energy
penalty for high-curvature surface areas and hence a spatial
distribution opposite to that of elastic density (see the dot-dashed curves).
This classical view of quantum dot formation has been well reproduced
in our results of all three evolution stages: the early morphological
instability shown in Fig. \ref{fig:muy_hy}(b) (at $t=1000$),
a coarsening regime in Fig. \ref{fig:muy_hy}(c) (at $t=2000$), and
a saturating stage in Fig. \ref{fig:muy_hy}(d)  (at $t=10000$). 

For islands to be saturated and stabilized, one would expect
mechanisms of film evolution involving additional energy penalty for
large, increasing island size, so that the overall stabilization
factors would compensate and suppress the destabilization effect 
(i.e., continuing growth and coarsening of surface islands) caused 
by stress relaxation. In previous studies such factors are usually
provided by additional surface energy terms particularly the surface 
energy anisotropy, which has been shown to enhance the surface-energy 
stabilization effect, constrain the island height, and lead to island 
shape/facet selection and transition \cite{re:ross98,re:ribeiro98};
this effect of surface anisotropy has been deemed essential 
for the existence of steady island arrays in some previous theoretical 
work (with various assumptions of the wetting effect) 
\cite{re:chiu99,re:eisenberg05,re:aqua10}. However, in this work we only 
consider isotropic surface energy. What we identify here is a new 
factor that is due to the contribution of higher-order perturbed
elastic energy on the interaction and evolution of surface islands, as
detailed in Fig. \ref{fig:muy_hy}: Positive contribution from the
2nd-order elastic energy density ${\cal E}^{(2)f}$ is found for large
surface islands, showing as an effective energy-penalty term and hence
a reduction of strain relaxation effect. [Note that this result is
still compatible with the well-known strain relaxation mechanism,
since the total elastic density ${\cal E}^{f}$ still shows a
destabilization effect due to the dominance of 1st-order density
${\cal E}^{(1)f}$; see Figs. \ref{fig:muy_hy} (c) and (d).] Such
effect of ${\cal E}^{(2)f}$ becomes important only at late stage with
large enough islands, and is negligible for small ones, as seen from
the comparison between Figs. \ref{fig:muy_hy} (b)-(d). 

To understand this seemingly counterintuitive result which is beyond 
the conventional view based on linear instability analysis, we
examine the detailed expression of ${\cal E}^{(2)f}$ which, from
Eqs. (\ref{eq:second})--(\ref{eq:lambda}), is rewritten as
\begin{equation}
{\cal E}^{(2)f} (\mathbf{r}) = {\epsilon^*}^2 \left [ f(h)
+ \int d \mathbf{r}' \int d \mathbf{r}'' h(\mathbf{r}')
G(\mathbf{r}-\mathbf{r}',\mathbf{r}'-\mathbf{r}'') 
h(\mathbf{r}'') \right ],
\label{eq:E_el2}
\end{equation}
where
\begin{equation}
f(h) = |\mathbf{\nabla} h|^2 + \nu \left ( {\cal E}^{(1)f}/{\epsilon^*}^2
\right )^2 + (1-\nu) \sum_{i=1}^{3} g_i^2(h),
\end{equation}
with $g_i(h)$ ($i=1,2,3$) the Fourier transform of $q_x^2 \hat{h}/q$,
$q_y^2 \hat{h}/q$, and $\sqrt{2} q_x q_y \hat{h}/q$ respectively, and
\begin{eqnarray}
&&G(\mathbf{r}-\mathbf{r}',\mathbf{r}'-\mathbf{r}'') =
\sum_{\mathbf{q},\mathbf{q}'} e^{i \mathbf{q} \cdot
  (\mathbf{r}-\mathbf{r}') + i \mathbf{q}' \cdot
  (\mathbf{r}'-\mathbf{r}'')} \nonumber \\
&& \times \frac{2}{qq'} \left \{ q_x(q_x-q_x')(q_x'^2+\nu q_y'^2)
  +q_y(q_y-q_y')(q_y'^2+\nu q_x'^2)+(1-\nu)q_x'q_y' \left [
    q_x(q_y-q_y')+q_y(q_x-q_x') \right ] \right \}.
\end{eqnarray}
In Eq. (\ref{eq:E_el2}), the first part $f(h)$ is always positive,
analogous to the ``self'' elastic energy of a given surface
profile that serves as a energy penalty to suppress its coarsening; 
the 2nd part represents the correlation between surface
heights and thus the elastic interaction between surface islands.
Within each island region (particularly near the peak), the magnitudes
of both parts increase with the island size as verified in our
numerical calculations.

\begin{figure}
\centerline{
\includegraphics[height=3.in]{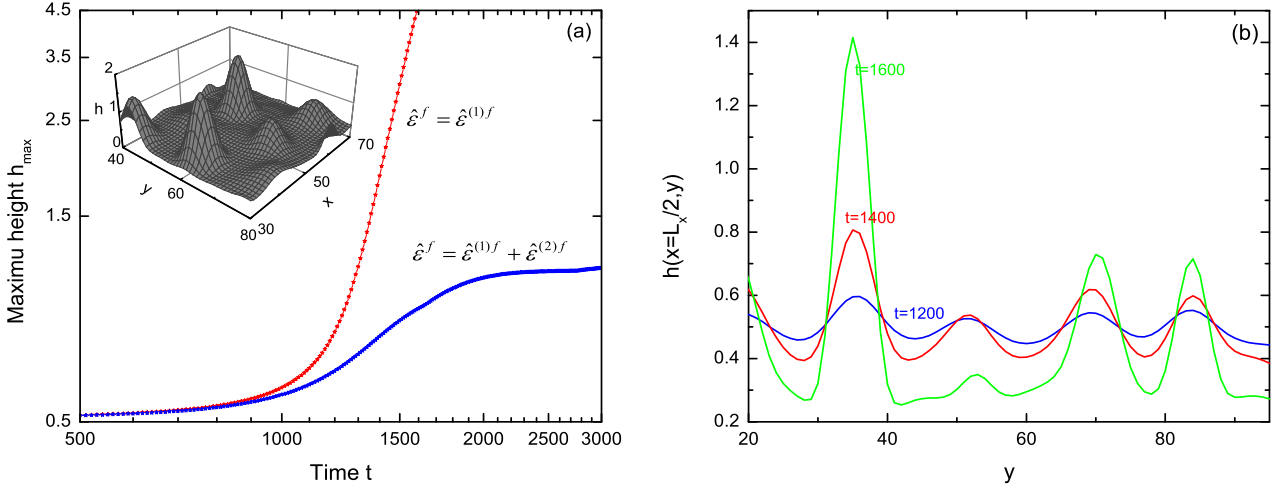}}
\caption{Results of strained film evolution with only first-order
elastic energy ${\cal E}^{(1)f}$ incorporated. All other parameters are 
the same as those in Figs. \ref{fig:muy} and \ref{fig:muy_hy}. 
(a) Maximum surface height as a function of time $t$, with results from 
calculations up to 2nd-order elastic energy (with ${\cal E}^f =
{\cal E}^{(1)f} + {\cal E}^{(2)f}$) also shown for comparison; a sample 
3D image of island morphology at $t=1600$ is presented in the inset.
(b) 2D cross-section profiles of film surface morphology at different
times.}
\label{fig:El1}
\end{figure}

If these 2nd-order elastic contributions are absent or not strong enough, 
the elastic energy relaxation would increasingly dominate over the
surface-energy stabilization effect, driving the continuing island 
growth even in the presence of the wetting potential.
This can be illustrated clearly from our numerical results given in
Fig. \ref{fig:El1}, where the same film evolution equation 
(\ref{eq:height}) is simulated, but with only first-order elastic
energy ${\cal E}^{(1)f}$ incorporated. All other parameters remain unchanged, 
including the same wetting potential approximation Eq. (\ref{eq:wetting}).
The maximum surface height is found to increase monotonically
with time [see Fig. \ref{fig:El1} (a)], without any slowing or saturation
process observed, a result that is consistent with previous work
\cite{re:levine07}. Time evolution of the corresponding 2D cross-section
surface profiles is given in Fig. \ref{fig:El1} (b), from which two main
features of surface dynamics can be identified: (1) Large mass transport
from film layers to islands is observed, leading to much thinner film
layers between surface islands as compared to the result shown in
Fig. \ref{fig:muy_hy} (a) which incorporates the 2nd-order elastic energy 
effects. Although the wetting potential still has the effect of preserving 
the wetting layer in-between surface islands and then limiting the diffusion 
process from the depleted wetting layer to the peaks, here such effect 
becomes relatively weaker as time evolves due to the increasing dominance 
of the destabilization effect of 1st-order elastic energy and the absence
of ``self'' energy penalty term $f(h)$ for large islands. 
(2) Mass transport between islands continues to occur, which corresponds
to island migration or coarsening process and is actually a secondary
effect compared to (1). This process cannot be prevented by the wetting
effect, and can be controlled only by the higher-order elastic energy 
terms describing island interaction and correlation [see Eq. (\ref{eq:E_el2})].
Thus at late times the island heights increase rapidly, resulting in the 
formation of surface islands with large aspect ratio between height and width 
as shown in Fig. \ref{fig:El1}. The perturbation method used here is no longer 
valid for such high islands, and the simulations will ultimately blow up. 
This is qualitatively different from the results given above with the 
incorporation of 2nd-order elastic energy, where islands with 
well constrained aspect ratio are obtained which also shows 
the applicability of the perturbation method developed here.

All these results indicate that the nonlinearities given by the higher-order 
strain energy of individual islands and the elastic interaction between 
islands can affect the pathway of film strain relaxation at late evolution 
times, slow down the decrease of total elastic energy via their increasing
positive energy contribution for large islands, and thus effectively reduce 
the effect of stress relaxation as the surface instability driving force. 
Such reduction leads to relatively stronger role played by the surface 
energy and the wetting potential at later times [see the comparison between 
Figs. \ref{fig:muy_hy} (c) and (d), and between Figs. \ref{fig:muy_hy} (a)
and \ref{fig:El1} (b)], limiting the mass transport between film layers and
islands and hence suppressing the island growth and coarsening.

%*************************************************************************
\section{Conclusions}

We have investigated the nonlinear dynamic processes governing the 
formation, coarsening, and stabilization of strained quantum dot 
islands on the surface of heteroepitaxial films, through the development 
of a nonlinear evolution equation for film morphology. Our study is based
on a continuum elasticity model that incorporates the film-substrate
wetting effect and importantly, on the construction of a perturbation
method in Fourier space for determining the system elastic properties.
In addition to a linear stability analysis which yields the conditions
of film morphological instability, we have performed large scale numerical 
calculations of the dynamic equation derived to study the detailed behavior 
of film evolution. We focus on effects of small misfit strains which
correspond to relatively large length scale of surface nanostructures, 
and analyze the evolution of strained surface islands/dots using a variety 
of characteristics of film morphology, including the structure factor of 
surface height, its first three moments, the maximum height of surface 
profile, and the surface roughness.

Consistent with previous experimental and theoretical work, 
our results have shown three characteristic
stages of island evolution for post-deposited annealing films, 
including (1) the early stage of morphological instability and 
island formation, as characterized by the exponential growth of 
maximum structure factor $S_{\rm max}$ and the surface roughness 
as well as the increase of maximum surface height and moments $m_n$;
(2) a nonlinear island coarsening stage, with a transient power-law
behavior of $m_n$ decay that appears at the beginning of
this stage; and (3) a crossover to an asymptotic state of saturated
island arrays (although without long-range spatial order), after the
slowing and suppression process of coarsening. Also,
the dependence of these detailed properties on the film-substrate
misfit strain has been obtained, such as the values of
coarsening exponents and the time ranges for the crossover between
different evolution stages. These have been shown important for the
understanding of different, or seemingly inconsistent, experimental
results particularly for the late time stage of island coarsening or
stabilization. On the other hand, such dependence does not qualitatively
affect our results of the three evolution regimes; same conclusion can
be drawn for the effect of different finite system sizes used in our
simulations. To understand the mechanisms underlying the nonlinear
evolution of strained films, we have examined the effects of 
film-substrate wetting potential, in particular its role on the
suppression of the valley-to-peak mass diffusion process that would lead 
to wetting layer depletion, and its constraining effect on island growth.
Furthermore, through a detailed study of time evolution of elastic energy
density distribution at the film surface, we find that higher-order terms 
of film elastic energy, which incorporate the interaction between
strained surface islands and the higher-order ``self'' elastic energy of 
individual islands, can effectively alter the relaxation pathway of film 
strain energy at late stage. They play an important role on the saturation 
and stabilization of quantum dot arrays, in particular the crossover to 
the saturated state with balanced multi-island interactions and limited
island-layer and between-island mass transport. Thus our results indicate 
that both effects of film-substrate wetting interaction and high-order 
elastic energy are pivotal for the achieving of steady quantum dot arrays 
and also for the understanding of self-assembly process of strained film 
heteroepitaxy.

\begin{acknowledgments}
This work was supported by the National Science Foundation
under Grant No. DMR-0845264.
\end{acknowledgments}

%*************************************************************************
\appendix*
\section{Third order perturbation results of film elasticity}

As described in Sec. \ref{sec:method}, our perturbation approach
developed here for solving the system elasticity problem can be
extended to obtain higher order results through a recursive procedure. 
We have calculated the elastic properties of this heteroepitaxial 
system up to third order, with results presented in this appendix. 
The 3rd-order perturbed elastic energy density is given by
\begin{eqnarray}
&\tilde{\mathcal E}^{(3)f}&=\frac{E\epsilon}{1-\nu} \left [
  (1-\nu)\frac{a_1^{(3)}q_x+b_1^{(3)}q_y} {\mu q}
  -(1-2\nu)\frac{c_1^{(3)}}{2\mu} \right ] \nonumber\\
&&+\sum_{\mathbf{q}'} \left \{ \frac{1+\nu}{2E} \left [
  \hat{\sigma}_{ij}^{(1)f}(\mathbf{q}')
  \hat{\sigma}_{ij}^{(2)f}(\mathbf{q}-\mathbf{q}') 
  +\hat{\sigma}_{ij}^{(2)f}(\mathbf{q}')
  \hat{\sigma}_{ij}^{(1)f}(\mathbf{q}-\mathbf{q}') \right ]
\right. \nonumber\\
&&\left. -\frac{\nu}{2E} \left [ \hat{\sigma}_{ll}^{(1)f}(\mathbf{q}')
    \hat{\sigma}_{ll}^{(2)f}(\mathbf{q}-\mathbf{q}')
    +\hat{\sigma}_{ll}^{(2)f}(\mathbf{q}')
    \hat{\sigma}_{ll}^{(1)f}(\mathbf{q}-\mathbf{q}') 
\right ] \right \},
\label{eq:energy3rd}
\end{eqnarray}
where
\begin{eqnarray}
&&a_1^{(3)}=\sum_{\mathbf{q}'} \left [
  (q_x-q_x') \hat{\sigma}_{xx}^{(2)f}(\mathbf{q}') 
  +(q_y-q_y') \hat{\sigma}_{xy}^{(2)f}(\mathbf{q}') \right ]
\hat{h}(\mathbf{q}-\mathbf{q}'),
\label{eq:a13rd}\\
&&b_1^{(3)}=\sum_{\mathbf{q}'} \left [
  (q_x-q_x')\hat{\sigma}_{xy}^{(2)f}(\mathbf{q}') 
  +(q_y-q_y')\hat{\sigma}_{yy}^{(2)f}(\mathbf{q}') \right ]
\hat{h}(\mathbf{q}-\mathbf{q}'),
\label{eq:b13rd}\\
&&c_1^{(3)}=\sum_{\mathbf{q}'}i \left [
  (q_x-q_x')\hat{\sigma}_{xz}^{(2)f}(\mathbf{q}')
  +(q_y-q_y')\hat{\sigma}_{yz}^{(2)f}(\mathbf{q}') \right ]
\hat{h}(\mathbf{q}-\mathbf{q}').
\label{eq:c13rd}
\end{eqnarray}

Based on the perturbation solutions of the system elasticity, we have
obtained the first and second order results for elastic stress
tensors at the film surface, which are used to calculate the perturbed
elastic energy density given above. In second order we have
\begin{eqnarray}
&\hat{\sigma}_{xx}^{(2)f}(\mathbf{q})&=-\frac{E\epsilon}{1-\nu}
\sum_{\mathbf{q}'} \left \{ 4\frac{q_x(q_x-q_x')}{q^3q'}
(\nu q_y'^2+q_x'^2)(q^2+\nu q_y^2)
+4\nu \frac{(q_y-q_y')q_y^3}{q^3q'}(q_y'^2+\nu q_x'^2)\right.\nonumber\\
&&+4(1-\nu)\frac{q_x'q_y'}{q^3q'}\left[q_x(q_y-q_y')+q_y(q_x-q_x')
\right ] (q_x^2+\nu q^2) -\frac{1}{q^2} \left [
  q_x'(q_x-q_x')+q_y'(q_y-q_y') \right] (2\nu q_y^2+q_x^2) \nonumber\\
&&\left.+4(1-\nu)\frac{q_xq_yq_x'q_y'}{q^3q'} \left [
    q_y(q_y-q_y')-q_x(q_x-q_x') \right] \right \}
\hat{h}(\mathbf{q}')\hat{h}(\mathbf{q}-\mathbf{q}'),
\label{eq:sigxx2}
\end{eqnarray}
\begin{eqnarray}
&\hat{\sigma}_{yy}^{(2)f}(\mathbf{q})&=-\frac{E\epsilon}{1-\nu} 
\sum_{\mathbf{q}'} \left \{ 4\frac{q_y(q_y-q_y')}{q^3q'}
(\nu q_x'^2+q_y'^2)(q^2+\nu q_x^2)
+4\nu \frac{(q_x-q_x')q_x^3}{q^3q'}(q_x'^2+\nu q_y'^2)\right.\nonumber\\
&&+4(1-\nu)\frac{q_x'q_y'}{q^3q'} \left [ q_x(q_y-q_y') +q_y(q_x-q_x')
\right ] (q_y^2+\nu q^2) -\frac{1}{q^2} \left [
q_x'(q_x-q_x')+q_y'(q_y-q_y') \right ] (2\nu q_x^2+q_y^2)\nonumber\\
&&\left.-4(1-\nu)\frac{q_xq_yq_x'q_y'}{q^3q'} \left [
    q_y(q_y-q_y')-q_x(q_x-q_x') \right ] \right \}
\hat{h}(\mathbf{q}')\hat{h}(\mathbf{q}-\mathbf{q}'),
\label{eq:sigyy2}
\end{eqnarray}
\begin{equation}
\hat{\sigma}_{zz}^{(2)f}(\mathbf{q})=\frac{E\epsilon}{(1-\nu)}(3-2\nu)
\sum_{\mathbf{q}'} \left [ q_x'(q_x-q_x')+q_y'(q_y-q_y') \right ]
\hat{h}(\mathbf{q}')\hat{h}(\mathbf{q}-\mathbf{q}'), 
\label{eq:sigzz2}
\end{equation}
\begin{eqnarray}
&\hat{\sigma}_{xy}^{(2)f}(\mathbf{q})&=E\epsilon\sum_{\mathbf{q}'}
\left\{-4\frac{q_xq_y}{q'q^3} \left[ \left ( 
   q_x(q_x-q_x')(q_x'^2+\nu q_y'^2) +q_y(q_y-q_y')(q_y'^2+\nu q_x'^2)
 \right ) \right. \right. \nonumber\\
&&\left.+(1-\nu)q_x'q_y' \left ( q_x(q_y-q_y')+q_y(q_x-q_x') \right )
\right ] +\frac{(1-2\nu)}{(1-\nu)} \frac{q_xq_y}{q^2} \left [ 
q_x'(q_x-q_x')+q_y'(q_y-q_y') \right ] \nonumber\\
&&-\frac{4}{(1-\nu)}\frac{q_y^2}{q^3q'} \left [
q_y(q_x-q_x')(q_x'^2+\nu q_y'^2)-q_x(q_y-q_y')(q_y'^2+\nu q_x'^2) 
\right.\nonumber\\
&&\left.\left.+(1-\nu) q_x'q_y' \left ( q_y(q_y-q_y')-q_x(q_x-q_x')
    \right ) \right ] \right \}
\hat{h}(\mathbf{q}')\hat{h}(\mathbf{q}-\mathbf{q}'),  
\label{eq:sigxy2}
\end{eqnarray}
\begin{eqnarray}
&\hat{\sigma}_{xz}^{(2)f}(\mathbf{q})&=\frac{2E\epsilon}{1-\nu}
\sum_{\mathbf{q}'} i\frac{q_x}{q^2q'} \left \{
q_x(q_x-q_x')(q_x'^2+\nu q_y'^2)+q_y(q_y-q_y')(q_y'^2+\nu q_x'^2) 
\right. \nonumber\\
&&+(1-\nu)q_x'q_y'\left [ q_x(q_y-q_y')+q_y(q_x-q_x') \right ]
-\frac{q_y^2}{q_x} \left [ (q_x-q_x')(q_x'^2+\nu q_y'^2)
+(1-\nu)(q_y-q_y')q_x'q_y' \right ] \nonumber\\
&&\left.+(1-\nu)q_y(q_x-q_x')q_x'q_y' 
+q_y(q_y-q_y')(q_y'^2+\nu q_x'^2) \right \}
\hat{h}(\mathbf{q}')\hat{h}(\mathbf{q}-\mathbf{q}'), 
\label{eq:sigxz2}
\end{eqnarray}
\begin{eqnarray}
&\hat{\sigma}_{yz}^{(2)f}(\mathbf{q})&=\frac{2E\epsilon}{1-\nu}
\sum_{\mathbf{q}'} i\frac{q_y}{q^2q'} \left \{
q_x(q_x-q_x')(q_x'^2+\nu q_y'^2)+q_y(q_y-q_y')(q_y'^2+\nu q_x'^2)
\right. \nonumber\\
&&+(1-\nu)q_x'q_y' \left [ q_x(q_y-q_y')+q_y(q_x-q_x') \right ]
-\frac{q_x^2}{q_y} \left [ (q_y-q_y')(q_y'^2+\nu q_x'^2)
+(1-\nu)(q_x-q_x')q_x'q_y'\right]\nonumber\\
&&\left.+(1-\nu)q_x(q_y-q_y')q_x'q_y' 
+q_x(q_x-q_x')(q_x'^2+\nu q_y^2) \right \}
\hat{h}(\mathbf{q}')\hat{h}(\mathbf{q}-\mathbf{q}'), 
\label{eq:sigyz2}
\end{eqnarray}

while the results for first order perturbation are given by
\begin{eqnarray}
&&\hat{\sigma}_{xx}^{(1)f}(\mathbf{q})
=\frac{2E\epsilon}{q(1-\nu)}(q_x^2+\nu q_y^2)\hat{h}(\mathbf{q}),
\label{eq:sigxx1} \\
&&\hat{\sigma}_{yy}^{(1)f}(\mathbf{q})
=\frac{2E\epsilon}{q(1-\nu)}(q_y^2+\nu q_x^2)\hat{h}(\mathbf{q}),
\label{eq:sigyy1} \\
&&\hat{\sigma}_{zz}^{(1)f}(\mathbf{q})=0,
\label{eq:sigzz1} \\
&&\hat{\sigma}_{xy}^{(1)f}(\mathbf{q})=2E\epsilon \frac{q_xq_y}{q}\hat{h}(\mathbf{q}),
\label{eq:sigxy1} \\
&&\hat{\sigma}_{xz}^{(1)f}(\mathbf{q})=-\frac{E\epsilon}{1-\nu}iq_x\hat{h}(\mathbf{q}),
\label{eq:sigxz1} \\
&&\hat{\sigma}_{yz}^{(1)f}(\mathbf{q})=-\frac{E\epsilon}{1-\nu}iq_y\hat{h}(\mathbf{q}).
\label{eq:sigyz1}
\end{eqnarray}

%*********************************************************************************

\end{document}